\documentclass[a4paper,UKenglish,utf8]{lipics}
\usepackage{subfig}

\newcommand{\cat}[1][C]{\mathbb{#1}}

\newcommand{\obcat}[1][\cat]{\mathrm{Obj}_{#1}}

\newcommand{\triskell}[2][\Omega]{\mathbb{T}\mathrm{k}^{\object{#1}}_{\simeq}(#2)}
\newcommand{\triskellnocong}[2][\Omega]{\mathbb{T}\mathrm{k}^{\object{#1}}(#2)}
\newcommand{\rel}[2][\Omega]{\mathbb{R}\mathrm{el}_{\object{#1}}(#2)}
\newcommand{\matcat}[1][\Omega]{\mathbb{M}\mathrm{at}_{\object{#1}}}

\newcommand{\set}[1][]{\mathbb{S}\mathrm{et}_{#1}}
\newcommand{\object}[1]{\mathrm{#1}}
\newcommand{\homset}[3][\cat]{\mathrm{Hom}_{#1}(\object{#2},\object{#3})}
\newcommand{\contract}[1][\cdot]{\mathcal{C}(#1)}

\newcommand{\alltriskells}[3]{\mathbf{T}\mathrm{k}^{\object{#3}}(\object{#1},\object{#2})}
\newcommand{\alltriskellssym}[2]{\alltriskells{#1}{#1}{#2}}

\newcommand{\fock}{\mathfrak{F}}
\newcommand{\symmfunctor}{\mathfrak{S}}
\newcommand{\liftedfock}{\mathfrak{F}_{\scriptsize{\uparrow}}}
\renewcommand{\contract}{\mathfrak{C}}
\newcommand{\mmap}{{\tt m}}
\newcommand{\umap}[1][]{{\tt u}_{\object{#1}}}
\newcommand{\trisk}[1][T]{\mathcal{#1}}

\newcommand{\prodmon}{\otimes}
\newcommand{\coprodmon}{\oplus}
\newcommand{\union}{\cup}

\newcommand{\finpow}{{\mathrm P}_{{\rm fin}}}
\newcommand{\mfin}{{\mathrm M}_{{\rm fin}}}

\newcommand{\cohorth}{\perp_\mathrm{coh}}
\newcommand{\goiorth}{\perp_\mathrm{goi}}

\newcommand{\qcs}{{\sc qcs}\xspace}
\newcommand{\IG}{{\sc ig}\xspace}

\usepackage{fullmacros-lipics}
\usetikzlibrary{calc}
{\normalfont}

\title{From Dynamic to Static Semantics, Quantitatively}
\author[1]{Thomas Seiller}
\affil[1]{Department of Computer Science, University of Copenhagen\\
  \texttt{seiller@di.ku.dk}}
\authorrunning{T. Seiller} %mandatory. First: Use abbreviated first/middle names. Second (only in severe cases): Use first author plus 'et. al.'

\Copyright{Thomas Seiller}%mandatory, please use full first names. LIPIcs license is "CC-BY";  http://creativecommons.org/licenses/by/3.0/

\subjclass{F.3.2 Semantics of Programming Languages}% mandatory: Please choose ACM 1998 classifications from http://www.acm.org/about/class/ccs98-html . E.g., cite as "F.1.1 Models of Computation". 
\keywords{Linear Logic, Dynamic Semantics, Quantitative Semantics}% mandatory: Please provide 1-5 keywords
\serieslogo{}%please provide filename (without suffix)
\volumeinfo%(easychair interface)
  {}% editors
  {0}% number of editors: 1, 2, ....
  {}% event
  {1}% volume
  {1}% issue
  {1}% starting page number
\EventShortName{}
\DOI{10.4230/LIPIcs.xxx.yyy.p}% to be completed by the volume editor
\begin{document}

\maketitle

\begin{abstract}
We exhibit a new relationship between dynamic and static semantics. We define the categorical outlay needed to define Interaction Graphs models, a generalisation of Girard's Geometry of Interaction models, which strongly relate to game semantics. We then show how this category is mapped to weighted relational models of linear logic. This brings into vision a new bridge between the dynamic and static approaches, and provides formal grounds for considering interaction graphs models as quantitative versions of GoI and game semantics models. We finally proceed to show how the interaction graphs models relate to a very general notion of quantitative coherence spaces.
\end{abstract}

\section{Introduction}

This paper is about denotational semantics of (a fragment of) linear logic. The focus is on (denotational) static and dynamic semantics, i.e. respectively \emph{semantics of proofs} and \emph{semantics of proofs and their cut-elimination}. Denotational semantics were introduced by Scott \cite{ScottOutline} as mathematical models for programming languages. Through the proofs-as-programs correspondence, it is equivalent to study semantics of programs and semantics of proofs, and the current paper will focus on a fragment of linear logic.

Denotational semantics are data-driven semantics, i.e. their focus is on data and data types. A program then corresponds to a function that maps some input data to some output data. This view of \emph{programs as functions} is however arguably too coarse, as it models only the \emph{what} (a program computes) and not the \emph{how} (the program computes). Consider a program $P$ and an input data $A$, and the result $B$ that $P$ outputs when given $A$ as input. Denotational semantics will represent both $P(A)$ -- $P$ given input $A$ -- and $B$ -- the result of $P$'s computation given input $A$ -- by the same object, i.e. $\Int{P(A)}{}=\Int{B}{}$. Dynamic semantics try to provide mathematical models that capture also \emph{how} programs compute. Formally a dynamic semantic will be such that $\Int{P(A)}{}\neq\Int{B}{}$, but will provide an interpretation to the computation as well, i.e. there exists an operation $\rightarrow$ in the model such that $\Int{P(A)}{}\rightarrow\Int{B}{}$. %In some ways, denotational semantics are therefore dynamic semantics were the interpretation of the computation -- the dynamics -- is trivial, i.e. $\rightarrow$ is taken as the equality. 

This paper presents a functor mapping the interpretations of Multiplicative Linear Logic (\MLL) proofs in dynamic semantics to the interpretations of \MLL proofs in denotational semantics. Intuitively, one can understand this functor as taking a dynamic semantics and collapsing the interpretation of execution to an equality. 
%In particular, we start with the author's Interaction graphs dynamic semantics which encompasses a very broad set of dynamic semantics, and show that this collapse can be performed in a way such that the result provides the usual denotational semantics. 
More precisely, we show how this collapse can be performed in a way that preserves \emph{quantitative information}. %and the collapse of Interaction Graphs models are shown to produce various so-called quantitative denotational semantics. 
Formally, the functor is defined as the antisymmetric tensor algebra construction.

Since the paper is concerned with semantics of a fragment of linear logic, we will now speak of proofs rather than programs and of cut-elimination rather than program execution. However, the reader should make a note to herself that, through the Curry-Howard correspondence, proof-theoretic and computer science terminologies are interchangeable.

%The terminology \emph{dynamic semantics} encompasses game semantics and geometry of interaction semantics, while the denotational models we will be considering will be generalisations of the relational model for linear logic as well as Girard's \emph{coherence spaces} semantics. 

\subparagraph{Static semantics.}
Static semantics are denotational models that \emph{do not account for} the dynamics, i.e.\ the interpretation of a proof in a static model should be an invariant of the cut-elimination. A well-studied static semantics for linear logic is the so-called \emph{relational model} where formulas $A,B$ are interpreted as sets $\Int{A}{},\Int{B}{}$ and proofs of the linear implication $A\multimap B$ are interpreted as binary relations on $\Int{A}{}\times\Int{B}{}$. Another example of static semantics considered in this paper is Girard's coherence spaces, which can be defined categorically as a tight double-glueing construction \cite{doubleglueing} applied to the relational model.

\emph{Quantitative} denotational semantics are those which capture quantifiable properties of
the programs it represents, such as time, space, and resource consumption.
That is, quantitative semantics are denotational models of computation that
mirror more information than the so-called \emph{qualitative} models, e.g. the relational model, coherence spaces. Origins of this approach can be traced back to Girard's work on functor models for lambda-calculus \cite{FoncteurAnalytiques} -- that inspired linear logic \cite{linearlogic} --  which exhibited for the first time a decomposition of the semantic interpretation of lambda-terms as Taylor series.

Quantitative semantics has also provided static semantics for various algebraic extensions of lambda calculus such as probabilistic lambda calculi \cite{probcoh}. By considering a refinement of the relational model, Laird \emph{et al.} \cite{quantdenot} provided a uniform account of several static models accounting for quantitative notions. Finally, some quantitative generalisations of coherence space semantics have been considered, namely probabilistic \cite{qcs,probcoh} and quantum coherence spaces \cite{qcs}. However, no uniform account of those has yet appeared in the literature.

\subparagraph{Dynamic Semantics.} We regroup under this terminology those semantics that do not solely account for proofs but also for their \emph{dynamics}, e.g. game semantics and various \emph{geometry of interaction} (GoI) constructions. Formally, the interpretation of a proof of $A\multimap B$ cut with a proof of $A$ will differ from the interpretation of the cut-free proof of $B$ obtained by cut-elimination. Although different models and techniques regrouped under this name may differ on certain aspects, their interpretation of programs/proofs is quite similar \cite{baillot-phd}. %Since this paper deals with multiplicative linear logic only, they turn out to be more than simply similar: they are the same .

On the technical level, we will work in this paper with the author's \emph{interaction graphs} construction. These models include all of Girard's geometry of interaction (GoI) models \cite{seiller-goiadd}, and provide a far-reaching generalisation of those. Indeed, interaction graphs should be understood as a quantitative variant of geometry of interaction. Moreover, since GoI models are strongly related to game semantics (even more so when we restrict to \MLL as it done in this paper), interaction graphs can be thought of as a quantitative generalisation of those as well. Interaction graphs thus provide a natural framework to deal with quantitative dynamic interpretations with great generality. 

One should note that GoI models differ from those of game semantics by how the formulas are interpreted. Game semantics start with interpretations of formulas (games) and then define the interpretations of proofs. On the contrary, and though proof interpretations are the same, GoI constructions define the interpretation of formulas from the \enquote{untyped} interpretation of proofs. Categorically, the latter construction is realised by a (tight) double-glueing construction \cite{doubleglueing} defined from a notion of orthogonality between morphisms.

\subparagraph{Contributions.}
The first contribution of this paper is the definition of the categorical foundations of the interaction graphs constructions through the introduction of \emph{triskells}. This categorical apparatus is of interest in itself as it generalises both the bicategory of spans and the categories of matrices and arrays over a semiring. Furthermore, unlike in the work of Laird \emph{et al.}, our definition is performed internally to the underlying category and does not refer to external objects. This opens the way to defining even wider generalisations based on the replacement of the underlying category of sets with e.g. any topos.

Secondly, we shed light onto a new relationship between dynamic and denotational semantics. This relationship is not only shown to hold for \emph{qualitative semantics}, but also for \emph{quantitative semantics}. This provides formal grounds to the claim that the author's interaction graphs construction should be understood as a quantitative version of dynamic semantics -- GoI and game semantics. Though not the first work bridging dynamic and static semantics \cite{haghverdiscottdenot,CalderonMcCusker}, the technique used here is novel and is the first to be applied to quantitative models.

Finally, we describe how this functor not only preserves the interpretation of proofs but also maps \emph{orthogonalities}. I.e.\ we show how this functor lifts to a map from the double-glueing constructions considered in interaction graphs and the double-glueing construction of coherence spaces. In particular, the specific model detailed in the first paper by the author \cite{seiller-goim} is shown to map to \enquote{strict} probabilistic coherence spaces. This correspondence, although confined to specific cases when considering categories of relations, can be lifted to a much more general correspondence when considering the generalisation of weighted relations provided by triskells. We end the paper by sketching a very general construction of \emph{quantitative coherence spaces} suggested by the previous sections.

%The contributions of this paper are the following:
%\begin{itemize}
%\item we show a new relationship between dynamic semantics, e.g.\ game semantics, and denotational semantics, e.g. relational semantics, in terms of a functor.
%\item we provide a categorical foundation for the author's Interaction graphs constructions;
%\item we show how Interaction Graphs relate to quantitative denotational models, providing formal grounds to the claim that IG are a quantitative generalisation of Girard's GoI models;
%\item we show how this functor lifts to the level of triskells, i.e. graphs, and how the triskell approach generalizes the quantitative relational model;
%\item we describe how this functor not only preserves the categorical structures needed to interpret proofs, but also maps orthogonalities, i.e. the basis for the DG construction -- obtaining GoI models or Coherent spaces -- in some specific situations.
%\end{itemize}

\subparagraph{Organisation of the paper.} The first section introduces the notion of \emph{triskell} in a category. Triskells are generalisations of spans that will account for the quantitative aspects available in the interaction graphs constructions. The category of triskells will then (Section \autoref{sec_weighted}) be shown to have the structure of a \emph{traced monoidal category}, i.e.\ it has the structure to define a GoI interpretation of proofs. We also show how triskells can be mapped to weighted relations in the sense of Laird \emph{et al.} \cite{quantdenot}. These two results show that the category of triskells can accomodate both quantitative GoI and quantitative static semantics. The monoidal products interpreting the linear logic tensor in these two kinds of models are not the same however. We thus introduce (\autoref{sec_fock}) an endofunctor which maps one monoidal product to the other. This functor thus maps the interpretation of a proof of \MLL in (quantitative) dynamic models to the interpretation of the same proof in (quantitative) denotational models.
Finally, we study in \autoref{sec_orthogonality} the double-glueing construction of GoI and show that it relates to the coherence space double-glueing through the functor just exhibited. This result is however quite specific as it is based upon the notion of determinant and trace of matrices. We thus end this paper by defining a notion of determinant and trace of triskells, allowing for the definition of both quantitative GoI models and quantitative coherence space semantics in a very general setting

%\noindent\textbf{Acknowledgments.} 

\subparagraph{Acknowledgements.} This work was partly supported by the European Commision's Marie Sk\l{}odowska-Curie Individual Fellowship (H2020-MSCA-IF-2014) 659920 - ReACT and the ANR 12 JS02 006 01 project \href{http://lipn.univ-paris13.fr/~pagani/pmwiki/pmwiki.php/Coquas/Coquas}{COQUAS}. The author also thanks Hugh Steele for his interest in this work, and his numerous comments about earlier versions of this paper. 

\section{Triskells}\label{sec_triskells}

The idea of triskells comes from the author's work on Interaction Graphs \cite{seiller-goim,seiller-goiadd}. Triskells are a generalisation of spans \cite{spans}. While the latter can be used to represent directed graphs, triskells possess the extra structure needed to interpret weighted edges.

\begin{definition}
A \emph{triskell} $\trisk[T]$ in a category $\cat$ is defined as three morphisms sharing the same source: it is a tuple $(\object{E},\object{S},\object{T},\object{\Omega},s,t,w)$ with $\object{E},\object{S},\object{T},\object{\Omega}\in\obcat, s\in\homset{E}{S}, t\in\homset{E}{T}, w\in\homset{E}{\Omega}$. We call $\object{S}$ (resp. $\object{T}$, resp. $\object{\Omega}$) and $s$ (resp. $t$, resp. $w$) the \emph{source object} (resp. \emph{target object}, resp. \emph{weight object}) and \emph{source map} (resp. \emph{target map}, resp. \emph{weight map}). We denote by $\alltriskells{S}{T}{\Omega}$ the set of all triskells with source object $\object{S}$, target object $\object{T}$ and weight object $\object{\Omega}$.
\end{definition} 

For composition to be defined, the set of weights must be fixed once and for all and endowed with an associative binary product, i.e.\ $\object{\Omega}$ should have the structure of a monoid.

\begin{definition}
Let $(\cat,\otimes,1)$ be a monoidal category with $\alpha,\rho,\lambda$ the associativity, right and left unit maps of the monoidal product. A monoid object $(\object{M}, \mmap, \umap)$ in $\cat$ is an object $\object{M}$ together with a multiplication $\mmap\in\homset{M\otimes M}{M}$ and a unit $\umap\in\homset{1}{M}$, which satisfy associativity, i.e.\ $\mmap\circ (\identity[\object{M}]\otimes \mmap)=\mmap\circ(\mmap\otimes\identity[\object{M}])\circ\alpha$, and identity, i.e.\ $\mmap\circ(\identity[\object{M}]\otimes \umap)=\rho_{\object{M}}$ and $\mmap\circ(\umap\otimes \identity[\object{M}])=\lambda_{\object{M}}$.
%\textcolor{red}{A semigroup $\object{\Omega}$ in a (finitely complete) category $\cat$ is a tuple $(\object{S},m)$ where $\object{S}$ is an object of $\cat$, a \emph{unit map} $\umap\in\homset{I}{$\object{\Omega}$}$ and a \emph{multiplication map} $\mmap\in\homset{$\Omega\times\Omega$}{$\object{\Omega}$}$, satisfying associativity, i.e.\ $\mmap\circ (\identity[\Omega]\times \mmap)=\mmap\circ(\mmap\times\identity[\Omega])\circ\alpha$ where $\alpha$ is the associativity map of the product, and identity, i.e.\ .}
\end{definition}

For our purposes, we will suppose that $\cat$ is a cartesian category, i.e.\ the monoidal product is the cartesian product and the unit is a terminal object. We will write $\times$ the product, and $[\cdot,\cdot]$ the associated operation of \emph{pairing} of morphisms. This hypothesis is enough to ensure the existence, for each object $\object{E}$ of a unit map $\umap[E]\in\homset{E}{\Omega}$, i.e.\ a map such that $\mmap\circ [f,\umap[E]]=f=\mmap\circ [\umap[E],f]$ for all $f\in\homset{E}{\Omega}$. %This is possible when the unit of the monoidal product is a terminal object, i.e.\ when the monoidal product is the cartesian product; we can define in this case $\umap[E]=\umap\circ !_{E}$. In the following, 
From now on, all monoid objects considered will be monoid objects for the cartesian monoidal structure over $\cat$.

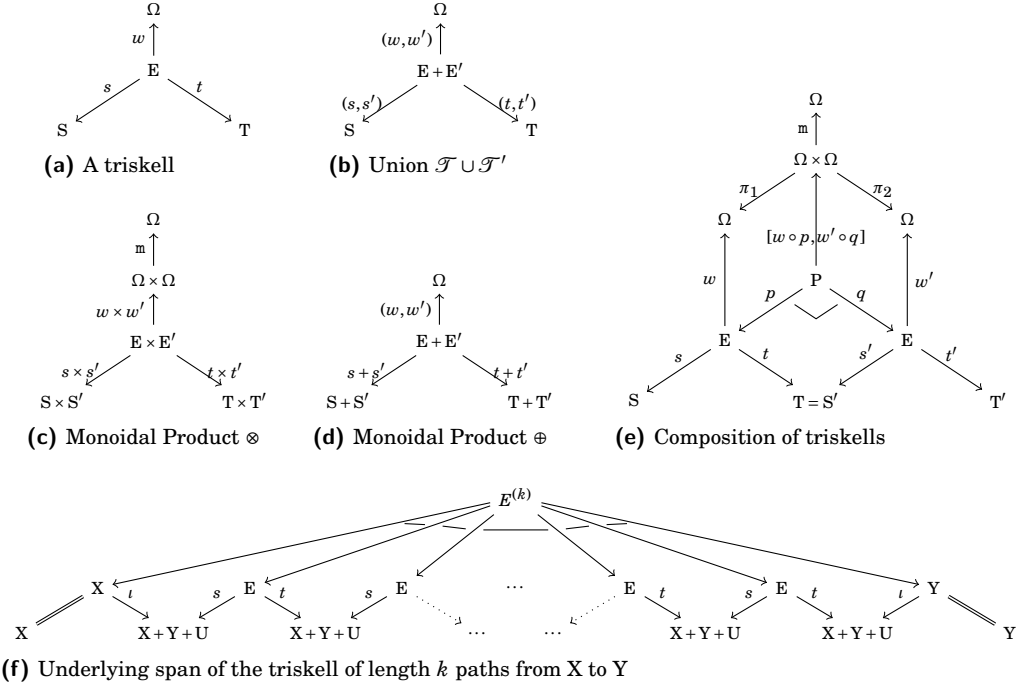
\begin{figure}[t]
\centering
\begin{tabular}{cc}
\begin{tabular}[b]{cc}
\subfloat[A triskell\label{Fig_Trisk}]{
\begin{tikzpicture}[x=0.6cm,y=0.4cm]
	\node (E1) at (-2,2) {\scriptsize{$\object{E}$}};
	\node (W1) at (-2,4) {\scriptsize{$\Omega$}};
	\node (S) at (-4,0) {\scriptsize{$\object{S}$}};
	\node (M) at (0,0) {\scriptsize{$\object{T}$}};
	
		\draw[->] (E1) -- (W1) node [midway,left] {\scriptsize{$w$}};
	\draw[->] (E1) -- (S) node [midway,above] {\scriptsize{$s$}};
	\draw[->] (E1) -- (M) node [midway,above] {\scriptsize{$t$}};
\end{tikzpicture}
}&
\subfloat[Union $\trisk\cup\trisk'$ \label{Fig_Trisk_Union}]{
\begin{tikzpicture}[x=0.6cm,y=0.4cm]
	\node (E1) at (-2,2) {\scriptsize{$\object{E} + \object{E'}$}};
	\node (W1) at (-2,4) {\scriptsize{$\Omega$}};
	\node (S) at (-4,0) {\scriptsize{$\object{S}$}};
	\node (M) at (0,0) {\scriptsize{$\object{T}$}};

	\draw[->] (E1) -- (W1) node [midway,left] {\scriptsize{$(w,w')$}};
	\draw[->] (E1) -- (S) node [midway,left] {\scriptsize{$(s,s')$}};
	\draw[->] (E1) -- (M) node [midway,right] {\scriptsize{$(t,t')$}};

\end{tikzpicture}
}\\
\subfloat[Monoidal Product $\prodmon$\label{Fig_Trisk_Prod}]{
\begin{tikzpicture}[x=0.6cm,y=0.4cm]
	\node (E1) at (-2,2) {\scriptsize{$\object{E}\times \object{E'}$}};
	\node (W1) at (-2,4) {\scriptsize{$\Omega\times \Omega$}};
	\node (W2) at (-2,6) {\scriptsize{$\Omega$}};
	\node (S) at (-4,0) {\scriptsize{$\object{S}\times \object{S'}$}};
	\node (M) at (0,0) {\scriptsize{$\object{T}\times \object{T'}$}};

	\draw[->] (E1) -- (W1) node [midway,left] {\scriptsize{$w\times w'$}};
	\draw[->] (W1) -- (W2) node [midway,left] {\scriptsize{$\mmap$}};
	\draw[->] (E1) -- (S) node [midway,left] {\scriptsize{$s\times s'$}};
	\draw[->] (E1) -- (M) node [midway,right] {\scriptsize{$t\times t'$}};

\end{tikzpicture}
}
&
\subfloat[Monoidal Product $\coprodmon$\label{Fig_Trisk_Sum}]{
\begin{tikzpicture}[x=0.6cm,y=0.4cm]
	\node (E1) at (-2,2) {\scriptsize{$\object{E}+\object{E'}$}};
	\node (W1) at (-2,4) {\scriptsize{$\Omega$}};
	\node (S) at (-4,0) {\scriptsize{$\object{S}+\object{S'}$}};
	\node (M) at (0,0) {\scriptsize{$\object{T}+\object{T'}$}};

	\draw[->] (E1) -- (W1) node [midway,left] {\scriptsize{$(w,w')$}};
	\draw[->] (E1) -- (S) node [midway,left] {\scriptsize{$s+s'$}};
	\draw[->] (E1) -- (M) node [midway,right] {\scriptsize{$t+t'$}};

\end{tikzpicture}
}
\end{tabular}
&
\subfloat[Composition of triskells\label{Fig_Trisk_Compo}]{
\begin{tikzpicture}[x=0.6cm,y=0.4cm]
	\node (E1) at (-2,2) {\scriptsize{$\object{E}$}};
	\node (W1) at (-2,6) {\scriptsize{$\Omega$}};
	\node (S) at (-4,0) {\scriptsize{$\object{S}$}};
	\node (M) at (0,0) {\scriptsize{$\object{T}=\object{S'}$}};
	\node (W2) at (2,6) {\scriptsize{$\Omega$}};
	\node (E2) at (2,2) {\scriptsize{$\object{E}$}};
	\node (T) at (4,0) {\scriptsize{$\object{T'}$}};
	\node (P) at (0,4) {\scriptsize{$\object{P}$}};
	\node (WW) at (0,8) {\scriptsize{$\Omega\times \Omega$}};
	\node (W3) at (0,10) {\scriptsize{$\Omega$}};

	\draw[->] (E1) -- (W1) node [midway,left] {\scriptsize{$w$}};
	\draw[->] (E1) -- (S) node [midway,above] {\scriptsize{$s$}};
	\draw[->] (E1) -- (M) node [midway,above] {\scriptsize{$t$}};
	\draw[->] (E2) -- (W2)  node [midway,right] {\scriptsize{$w'$}};
	\draw[->] (E2) -- (M) node [midway,above] {\scriptsize{$s'$}};
	\draw[->] (E2) -- (T) node [midway,above] {\scriptsize{$t'$}};
	
	\draw[->] (P) -- (E1)  node [midway,above] {\scriptsize{$p$}} node (P1) [near start] {};
	\draw[->] (P) -- (E2)  node [midway,above] {\scriptsize{$q$}} node (P2) [near start] {};
	\draw[-] (P1) -- ($(P1)+(P2)-(P)$) -- (P2) {};
	\draw[->] (WW) -- (W1) node [midway,above,left] {\scriptsize{$\pi_{1}$}};
	\draw[->] (WW) -- (W2) node [midway,above,right] {\scriptsize{$\pi_{2}$}};
	\draw[->] (P) -- (WW) node [midway,below] {\scriptsize{$[w\circ p,w'\circ q]$}};
	\draw[->] (WW) -- (W3) node [midway,left] {\scriptsize{$\mmap$}};
\end{tikzpicture}
}
\end{tabular}
\\
\subfloat[Underlying span of the triskell of length $k$ paths from $\object{X}$ to $\object{Y}$\label{Fig_Trisk_Paths}]{
\begin{tikzpicture}[x=0.5cm,y=0.3cm]
	\node (E1) at (-3,2) {\scriptsize{$\object{X}$}};
%	\node (W1) at (-2,6) {\scriptsize{$W$}};
	\node (S) at (-5,0) {\scriptsize{$\object{X}$}};
	\node (M) at (-1,0) {\scriptsize{$\object{X}+\object{Y}+\object{U}$}};
%	\node (W2) at (2,6) {\scriptsize{$W$}};
	\node (E2) at (1,2) {\scriptsize{$\object{E}$}};
	\node (T) at (3,0) {\scriptsize{$\object{X}+\object{Y}+\object{U}$}};
	
	\node (E3) at (5,2) {\scriptsize{$\object{E}$}};
	\node (T2) at (7,0) {\scriptsize{$\dots$}};	
	\node (Edots) at (8,2) {\scriptsize{$\dots$}};	
	\node (T22) at (9,0) {\scriptsize{$\dots$}};	
	\node (E4) at (11,2) {\scriptsize{$\object{E}$}};
	\node (T3) at (13,0) {\scriptsize{$\object{X}+\object{Y}+\object{U}$}};
	\node (E5) at (15,2) {\scriptsize{$\object{E}$}};
	\node (T4) at (17,0) {\scriptsize{$\object{X}+\object{Y}+\object{U}$}};	
	\node (E6) at (19,2) {\scriptsize{$\object{Y}$}};
	\node (T5) at (21,0) {\scriptsize{$\object{Y}$}};	
	
	\node (P) at (8,6) {\scriptsize{$E^{(k)}$}};
		\draw[->] (P) -- (E1) node [near start] (A1) {};
		\draw[->] (P) -- (E2) node [near start] (A2) {};
		\draw[->] (P) -- (E3) node [near start] (A3) {};
		\draw[->] (P) -- (E4) node [near start] (A4) {};
		\draw[->] (P) -- (E5) node [near start] (A5) {};
		\draw[->] (P) -- (E6) node [near start] (A6) {};
		\draw[-] (A1) -- (A2)-- (A3)-- (A4)-- (A5)-- (A6) {};
		
%	\node (WW) at (0,8) {\scriptsize{$W\times W$}};
%	\node (W3) at (0,10) {\scriptsize{$W$}};

%	\draw[->] (E1) -- (W1) node [midway,left] {\scriptsize{$w$}};
	\draw[-,double] (E1) -- (S) {};
	\draw[->] (E1) -- (M) node [midway,above] {\scriptsize{$\iota$}};
%	\draw[->] (E2) -- (W2)  node [midway,right] {\scriptsize{$w$}};
	\draw[->] (E2) -- (M) node [midway,above] {\scriptsize{$s$}};
	\draw[->] (E2) -- (T) node [midway,above] {\scriptsize{$t$}};
	
	\draw[->] (E3) -- (T) node [midway,above] {\scriptsize{$s$}};
	\draw[->,dotted] (E3) -- (T2) {};
	
	\draw[->,dotted] (E4) -- (T22) {};
	\draw[->] (E4) -- (T3) node [midway,above] {\scriptsize{$t$}};
	
	\draw[->] (E5) -- (T3) node [midway,above] {\scriptsize{$s$}};
	\draw[->] (E5) -- (T4) node [midway,above] {\scriptsize{$t$}};
	\draw[->] (E6) -- (T4) node [midway,above] {\scriptsize{$\iota$}};
	\draw[-,double] (E6) -- (T5);
	
%	\draw[->] (P) -- (E1)  node [midway,above] {\scriptsize{$p$}} node (P1) [near start] {};
%	\draw[->] (P) -- (E2)  node [midway,above] {\scriptsize{$q$}} node (P2) [near start] {};
%	\draw[-] (P1) -- (0,2.75) -- (P2) {};
%	\draw[->] (WW) -- (W1) node [midway,above,left] {\scriptsize{$\pi_{1}$}};
%	\draw[->] (WW) -- (W2) node [midway,above,right] {\scriptsize{$\pi_{2}$}};
%	\draw[->] (P) -- (WW) node [midway,below] {\scriptsize{$[w\circ p,w'\circ q]$}};
%	\draw[->] (WW) -- (W3) node [midway,left] {\scriptsize{$\mmap$}};
\end{tikzpicture}
}
\caption{Operations on triskells\label{Fig_Trisk_All}}
\end{figure}

We now fix a cartesian category $\cat$ with pullbacks, i.e.\ $\cat$ is finitely complete. We also pick a monoid object $\object{\Omega}$, and two triskells $\trisk[T],\trisk[T']$ with weight object $\object{\Omega}$ and such that the source object of $\trisk[T']$ coincides with the target object of $\trisk[T]$. We can define (\autoref{Fig_Trisk_Compo}) their \emph{composition} $\trisk[T']\circ \trisk[T]$ as the tuple $(\object{P},\object{S},\object{T},\object{W},q\circ s,p\circ t,\mmap\circ [w\circ p,w'\circ q])$, where $(\object{P},p,q)$ is a designated pullback of $t$ and $s'$. As when working with spans, associativity is only satisfied up to isomorphism and the natural structure we obtain is not that of a category but rather a bicategory. However, identifying isomorphic objects leads to a category.

\begin{definition}
Let $\cat$ be a complete category and $\object{\Omega}$ a monoid in $\cat$. The bicategory $\triskellnocong{\cat}$ is:
$$
\begin{array}{rcl}
\obcat[\triskellnocong{\cat}]=\obcat,
&\hspace{2cm}&
\homset[\triskellnocong{\cat}]{A}{B}=\alltriskells{A}{B}{\Omega}.
\end{array}
$$
The category $\triskell{\cat}$ is then defined as the quotient of $\triskellnocong{\cat}$ w.r.t. 2-isormorphisms.
\end{definition}

\begin{notation}
When they exist, coproducts will be denoted by $+$, while the \emph{copairing} of morphisms will be denoted by $(\cdot,\cdot)$. %: if $f\in\homset{X}{Z}$ and $g\in\homset{Y}{Z}$, $(f,g)\in\homset{X+Y}{Z}$. 
Given $f\in\homset{X}{Y}$ and $g\in\homset{X'}{Y'}$, we also define the morphism $f+g\in\homset{X+X'}{Y+Y'}$ as the copairing $(\iota_{1}\circ f,\iota_{2}\circ g)$ where $\iota_{1}\in\homset{Y}{Y+Y'}$ and $\iota_{2}\in\homset{Y'}{Y+Y'}$ are the natural inclusions.
\end{notation}

The product of the underlying category $\cat$ naturally induces a monoidal product on triskells written $\prodmon$ (\autoref{Fig_Trisk_Prod}). If $\cat$ has coproducts, then $\triskell{\cat}$ has two additional operations: the sum (\autoref{Fig_Trisk_Sum}) another monoidal product written $\coprodmon$, and the union, written $\union$ (\autoref{Fig_Trisk_Union}), which is defined between two triskells with same source and target objects.

\begin{notation}
We will denote by $1=(\{\star\},\{\star\},\{\star\},\Omega,!_{\{\star\}},!_{\{\star\}},\umap[\{\star\}])$ -- where $!_{\{\star\}}$ is the unique morphism $\{\star\}\rightarrow\{\star\}$ -- the unit of the monoidal product $\prodmon$. Whenever the underlying category has coproducts, we will denote by $\emptyset=(\emptyset,\emptyset,\emptyset,\Omega,\emptyset_{\emptyset},\emptyset_{\emptyset},\emptyset_{\Omega})$ -- where $\emptyset_{A}$ denotes the unique map $\emptyset\rightarrow A$ -- the unit of the monoidal product $\coprodmon$.
\end{notation}

The category of triskells is of interest since it is the ambient category in which the author's Interaction Graphs construction takes place \cite{seiller-goim} (taking $\cat$ as the category of countable sets and restricting to finite objects). Although later works \cite{seiller-goiadd,seiller-goig} make use of a larger category (in particular, one considers a sort of abelian enrichment to deal with additives \cite{seiller-goiadd}), we restrict our discussion to the category of triskells in this paper. 

\begin{example}
If $\cat$ is the category of countable sets $\set[\aleph_{1}]$, then a triskell over $\object{\Omega}$ whose source and target objects are finite sets is exactly a $\object{\Omega}$-weighted directed graph\footnote{Notice that in the presence of corpoducts, the notion of span captures exactly that of \emph{directed graphs}, which are usually represented by a pair of morphisms $s,t$ from a set of edges $\object{E}$ to a set of vertices $\object{V}$.\label{remgraphs}} as defined in earlier work by the author \cite{seiller-goim}. Composition corresponds to the composition of graphs by considering length 2 paths, while the trace defined below coincides with the \emph{execution} \cite{seiller-goim}.
\end{example}

The previous remarks explain the following terminology. When considering a triskell $(\object{E},\object{S},\object{T},\object{\Omega},s,t,w)$ over the category of sets, we will refer to the set $\object{E}$ as the \emph{set of edges}, and we will refer to any element of $\object{E}$ as an \emph{edge}.

\section{Triskells over $\set$ and weighted relations}\label{sec_weighted}

We now fix once and for all the underlying $\cat$ to be a category of sets, i.e. $\cat$ is the category $\set[\kappa]$ of all sets of cardinality less than $\kappa$ where $\kappa$ is any fixed cardinal. %Notice $\set[\kappa]$ is not necessarily cartesian closed.
 We suppose moreover that $\set[\kappa]$ has countable coproducts, i.e. $\kappa\geqslant\aleph_{1}$.

\subsection{Traced monoidal structure} 

We now explain how one can define a categorical trace \cite{tmc} on $\triskell{\cat}$ by constructing the \emph{triskell of paths}. We start from a triskell $\trisk[T]$ whose source and target objects are of the form $\object{X}+\object{U}$ and $\object{Y}+\object{U}$ respectively. One can define from it a triskell with same source and target $\object{X}+\object{Y}+\object{U}$ which we denote $\trisk'$. The composition of $k$ copies of this triskell $\trisk'$ with the same source and target objects $\object{X}+\object{Y}+\object{U}$ defines the \enquote{triskell of paths of length $k$}. Pre- and post-composition with \enquote{projections} onto $\object{Y}$ and $\object{X}$ respectively gives rise to the triskell of paths of length $k$ from $\object{X}$ to $\object{Y}$ (\autoref{Fig_Trisk_Paths}). Finally, countable coproducts allows to build the triskell $\Tr_{\object{U}}(\trisk[T])$ of all (finite) paths from $\object{X}$ to $\object{Y}$ in $\trisk[T]$, through the countable version of the union of triskells (\autoref{Fig_Trisk_Union}).
% and pre- and post-compositions with the inclusions of $\object{X}$ and $\object{Y}$ respectively in $\object{X}+\object{Y}$. %One then checks this defines %\footnote{Notice that a trace should be defined also when the morphism $A$ has different source and target, namely $X+U$ and $Y+U$ respectively. This is easily done by using injections of $X+U$ and $Y+U$ into $X+Y+U$.}
%a categorical trace . 

\begin{theorem}\label{thm_tracedmonoidal}
For any monoid $\object{\Omega}$ in $\set[\kappa]$, $\triskell{\set[\kappa]}$ is traced monoidal.
\end{theorem}

\begin{remark}
The above theorem should hold in a more general setting. First, the construction of the trace can be performed as long as the underlying category $\cat$ is finitely complete and has countable coproducts. Then, the proof that $\triskell{\cat}$ is indeed traced monoidal relies mainly on two additional properties of the category of sets, namely that coproducts are disjoint and stable under pullbacks. In particular, one should be able to replace $\set[\kappa]$ by any topos.
\end{remark}

%\section{Semirings and the Contracting Functor}

%We now fix the ambiant category $\cat$ as the category $\set$. Notice that one could fix $\kappa$ a cardinal and consider the category $\set[\kappa]$ of all sets below $\kappa$.

%\begin{proposition}
%The disjoint union of sets induces a monoidal structure on $\triskell{\cat}$. With this monoidal product, $\triskell{\cat}$ is a traced monoidal category. 
%\end{proposition}

%\textcolor{red}{Is this a coproduct or not?}

%\begin{proposition}
%The product of sets induces a monoidal structure on $\triskell{\cat}$.
%\end{proposition}

%\textcolor{red}{Is this a product or not?}

\subsection{Simple triskells and weighted relations}

When $\object{\Omega}$ is a complete semiring, one can define a functor from the category of triskells over $\object{\Omega}$ to the category of weighted relations over $\object{\Omega}$ through the following \emph{contraction of triskells}.

Given a triskell $T=(\object{E},\object{S},\object{T},\object{\Omega},s,t,w)$, the map $[s,t]$ factors uniquely as $\object{E}\stackrel{e}{\rightarrow} \object{\bar{E}}\stackrel{m}{\rightarrow} \object{S}\times \object{T}$, where $m$ is a monomorphism (and $e$ is an epimorphism). We can additionally define a map $\bar{w}:\object{\bar{E}}\rightarrow \object{W}$ as follows: $\bar{w}(a)=\sum_{b\in \object{E}, e(b)=a} w(b)$. We can therefore defined a triskell $\what{\trisk[T]}$ -- the contraction of $\trisk[T]$ -- as $(\object{\bar{E}},\object{S},\object{T},\object{W},\pi_{1}\circ m,\pi_{2}\circ m,\bar{w})$. This triskell has a particular property: the pairing of the source and target maps is a monomorphism; such triskells will be called \emph{simple}.

%\begin{center}
%\begin{tikzpicture}[x=0.4cm,y=0.4cm]
%	\node (E) at (2,2) {\scriptsize{$E_{M}$}};
%	\node (A) at (0,0) {\scriptsize{$A$}};
%	\node (B) at (4,0) {\scriptsize{$B$}};
%	\node (F) at (6,2) {\scriptsize{$E_{N}$}};
%	\node (C) at (8,0) {\scriptsize{$C$}};
%	\node (EF) at (4,4) {\scriptsize{$E_{M}\circ E_{N}$}};
%	
%	\draw[->] (E) -- (A) node [midway,left] {\scriptsize{$s$}};
%	\draw[->] (E) -- (B) {};
%	\draw[->] (F) -- (B) {};
%	\draw[->] (F) -- (C) node [midway,right] {\scriptsize{$t$}};
%	\draw[->] (EF) -- (E) node [near start] (N1) {} node [midway,left] {\scriptsize{$p$}};
%	\draw[->] (EF) -- (F) node [near start] (N2) {} node [midway,right] {\scriptsize{$q$}};
%	\draw[-] (N1) -- ($(N1)+(N2)-(EF)$) -- (N2) {};
%
%	\node (EF2) at (14,4) {\scriptsize{$E_{M}\circ E_{N}$}};
%	\node (A2) at (11,0) {\scriptsize{$A$}};
%	\node (B2) at (17,0) {\scriptsize{$C$}};
%	\node (AB) at (14,0) {\scriptsize{$A\times C$}};
%	\node (EF3) at (14,2) {\scriptsize{$E_{M}\bullet E_{N}$}};
%	
%	\draw[->] (EF2) -- (A2) node [midway,left] {\scriptsize{$p\circ s$}};
%	\draw[->] (EF2) -- (B2) node [midway,right] {\scriptsize{$q\circ t$}};
%	\draw[->>] (EF2) -- (EF3) node [midway,right] {\scriptsize{$e$}};
%	\draw[->] (AB) -- (A2) node [midway,below] {\scriptsize{$\pi_{1}$}};
%	\draw[->] (AB) -- (B2) node [midway,below] {\scriptsize{$\pi_{2}$}};
%	\draw[>->] (EF3) -- (AB) node [midway,right] {\scriptsize{$m$}};
%	\draw[->] (EF3) -- (A2) {};
%	\draw[->] (EF3) -- (B2) {};
%\end{tikzpicture}
%\end{center}

\begin{remark}
This mimics a natural contracting operation \cite{seiller-goim} that maps a (weighted) graph $G=(V,E,s,t,w)$ to a simple (weighted) graph $\what{G}$ with the same set of vertices $V$. %Given $s,t\in V$, let us write $G[s,t]$ the set of such edges in $G$. We define a single edge $(s,t)$ of source $s$ and target $t$ in $\what{G}$ if and only if $G[s,t]$ is non empty. This edge can be given a natural weight by summing the weights $w(e)$ for all $e\in G[s,t]$. I.e. $$\what{G}=(V,\{(s,t)\in V\times V'~|~ G[s,t]\neq\emptyset\},\pi_{V},\pi_{V'},(s,t)\mapsto \sum_{e\in G[s,t]}w(e))$$
\end{remark}

%\begin{definition}
%Let $\cat$ be a ... category. The category $\rel{\cat}$ is the category of \emph{simple triskells}, i.e.\ triskells such that the pairing $[s,t]$ is a monomorphism. Composition is defined as follows: the triskell composition defines a (generally) non-simple triskell $(\object{P},\object{S},\object{T},\object{W},s,t,w)$. Then the map $[s,t]$ factors as $P\stackrel{e}{\rightarrow} E\stackrel{m}{\rightarrow} S\times T$, where $m$ is a monomorphism. We can additionally define a map $\bar{w}:E\rightarrow W$ as follows: .... The composition of triskells is then defined as the simple triskell $(\object{E},\object{S},\object{T},\object{W},\pi_{1}\circ m,\pi_{2}\circ m,\bar{w})$.
%\end{definition}
%
%The following proposition relates this definition with the categories of \emph{weighted relations} defined by Laird and al. \cite{quantdenot}.

This operation on triskells actually defines a functor from the category of triskells to the category of weighted relations defined by Laird \emph{et al.} \cite{quantdenot} -- or equivalently the category $\matcat$ of matrices with coefficients in a (complete) semiring $\object{\Omega}$.

\begin{definition}
Let $\object{\Omega}$ be a complete semiring. We define the category $\rel{\set}$ as follows. Objects are sets and morphisms from $\object{A}$ to $\object{B}$ are matrices in $\object{\Omega}^{\object{A\times B}}$. Composition, written $\bullet$, is defined as the matrix product. The identity morphism is the identity matrix.
\end{definition}

Now, a matrix $M$ in $\object{\Omega}^{\object{A\times B}}$ can be written as a triskell $\trisk[T]_{M}$: writing $w:(a,b)\mapsto m_{a,b}$, we define this triskell by $\trisk[T]_{M}=(\object{A\times B},\object{A},\object{B},\object{\Omega},\pi_{1},\pi_{2},w)$. This correspondence is not one-to-one however, as zero coefficients can be represented both by edges with weight zero (as in the triskell defined above) or by a lack of edge. We therefore define also the minimal triskell associated to $M$ as $(\object{E}_{M},\object{A},\object{B},\object{\Omega},\pi_{1},\pi_{2},\bar{w})$ where $\object{E}_{M}=\{(a,b)\in \object{A}\times \object{B}~|~ m_{a,b}\neq 0\}$ and $\bar{w}$ is the restriction of $w$ to $\object{E}_{M}$. Conversely, every simple triskell $T=(\object{E},\object{A},\object{B},\object{\Omega},s,t,w)$ defines a matrix $M_{\trisk[T]}$ in $\object{\Omega}^{\object{A}\times \object{B}}$. Since $[s,t]$ is a monomorphism, we can consider $\object{E}$ a subset of $\object{S}\times \object{T}$. Then the coefficient $m_{a,b}$ is equal to $w((a,b))$ if $(a,b)\in \object{E}$ and null otherwise.

\begin{remark}
To obtain an exact correspondence, one should consider relations weighted in a complete semiring $\object{\Omega}$ (with a zero) and simple triskells over $\object{\Omega^{\ast}}$ -- $\object{\Omega}$ bereft of its zero element.
\end{remark}

It should be noted that the composition of relations is \emph{not} the composition of (simple) triskells: the composition $\trisk[T]\circ \trisk[T']$ of two simple triskells $\trisk[T],\trisk[T']$ need not be simple in general\footnote{This is already true for spans. To see this, it is enough to consider any triskells whose underlying spans are $(\{(1,2),(1,3)\},\{1\},\{2,3\},\pi_1,\pi_2)$ and $(\{(2,4),(3,4)\},\{2,3\},\{4\},\pi_1,\pi_2)$, with $\pi_i$ the natural projections.}. However, a simple computation shows that $\what{\trisk[T]_{M}\circ \trisk[T]_{N}}=\trisk[T]_{N\bullet M}$ for all matrices $M,N$ in $\object{\Omega}^{\object{A}\times\object{B}}$.

\begin{remark}
Two monoidal products are considered by Laird \emph{et al.} on the category $\rel{\set}$, namely the \emph{sum} $\oplus$ and the \emph{tensor} $\otimes$. It is not difficult to check that, through the identification of relations with simple triskells, those correspond to the monoidal products $\coprodmon$ and $\prodmon$ on triskells. Furthermore, the operation of union of triskells corresponds to matrix addition.
\end{remark}

%However, one can check that composition of relation is obtained by composing as triskells first and then using the contracting operation $\contract$, i.e.\ $T\bullet T'= \contract(T\circ T')$. This remark is key to the fact that the contracting operation defines a functor.

%To define correctly the contracting functor, one needs to use a factorisation system, i.e.\ one should be able to factorise every morphism in the category as a morphism followed by a monomorphism. This is easily done if one does not require any property for the first map, following Bousfield's construction of \enquote{cofibrantly generated} factorisation systems. Take the set of monomorphisms $M$ and then construct the factorisation system $({}^{\bot}M,({}^{\bot}M)^{\bot})$. Now, we want that $({}^{\bot}M)^{\bot}=M$. \textcolor{red}{(see cofibrantly generated model categories in nlab) It seems that $({}^{\bot}M)^{\bot}$ is equal to $\textnormal{cof}(M)$, the class of retracts in $\textnormal{cell}(M)$, which is itself the class of transfinite compositions of pushouts of elements of $M$. See Effective monomorphisms, which are closed under retracts. Are effective monomorphisms closed under transfinite compositions of pushouts?}

\begin{definition}
Let $\object{\Omega}$ be a complete semiring. We define the contraction functor $\contract$ from $\triskell{\cat}$ to $\rel{\cat}$ as follows. It acts as the identity on objects, and maps a triskell $\trisk[T]$ to the matrix $M_{\what{\trisk[T]}}$ of the contracted triskell $\what{\trisk[T]}$.
%Given a triskell $(\object{E},\object{S},\object{T},\object{W},s,t,w)$, we define a simple triskell $(\object{\bar{E}},\object{S},\object{T},\object{W},\bar{s},\bar{t},\bar{w})$ where $\bar{E}$ is the quotient of $E$ by the equivalence relation $e\sim e' \Leftrightarrow s(e)=s(e')\wedge t(e)=t(e')$. Then $\bar{s}$ and $\bar{t}$ are the natural induced source and target maps, and $\bar{w}$ is defined as $\bar{w}(\bar{e})=\sum_{e\in\bar{e}}w(e)$.
%We factorise the map $[s,t]$ as a map $g$ followed by a monomorphism $m$. The middle object $\bar{E}$ maps to $S$ and $T$ through $m\pi_{1}$ and $m\pi_{2}$, defining a simple span. We now have to take care of the weight map. For each point $a:\star\rightarrow \bar{E}$ we define the pushout $E_{a}$ of $a$ and $g$. If we have an exponential object we obtain a map from $\star$ to $\Omega^{E_{a}}$, which maps to $\object{\Omega}$ using the addition map $§:\Omega\times\Omega\rightarrow\Omega$ and completeness.
\end{definition}

%\begin{remark}
%It should be noticed here that instead of assuming that $\Omega$ already possess a complete semiring structure, one could have very well defined the functor $\contract$ from $\triskell{\cat}$ to $\rel[\mathcal{S}\Omega]{\cat}$ where $\mathcal{S}\Omega$ denotes the free complete semiring over $\Omega$. Both approaches are however equivalent, and we chose 
%\end{remark}

%\begin{theorem}
%There is an equivalence of categories between $\triskellsimp{\set}$ and the category of $\object{\Omega}$-weighted relations.
%\end{theorem}

%However, the contracting functor is not monoidal in the sense that the usual monoidal product on the category of weighted relations is not defined from the disjoint union.

\begin{theorem}\label{thm_contract_monoidal}
The functor $\contract$ is a strict monoidal functor for the two monoidal structures, i.e. $\contract(\trisk[T]\prodmon \trisk[T'])=\contract(\trisk[T])\prodmon \contract(\trisk[T'])$ and $\contract(\trisk[T]\coprodmon \trisk[T'])=\contract(\trisk[T])\coprodmon \contract(\trisk[T'])$.
\end{theorem}

This suggests the category $\triskell{\set}$ as a natural generalisation of the category of weighted relations. We will try to follow this intuition in the next sections. 

We have now shown that triskells can be used to define both quantitative denotational and dynamic semantics for \MLL. However, these interpretations are based upon the interpretation of the tensor product as two distinct monoidal products. In the next section, we explain how one can define a strict monoidal functor on the category of triskells which maps the monoidal product $\coprodmon$ onto the monoidal product $\prodmon$.

%\begin{proof}
%Although it is already called a functor, we first need to check that we defined a functor indeed. We only need to check that composition is respected. Taking two triskells $T$ and $T'$, we will consider the simple triskells $\contract(T\circ T')$ and $\contract(T)\bullet\contract(T')$. The fact that the underlying spans coincide is easily checked. The coicindence of weight maps is a consequence of the distributivity of the product over the sum in $\object{\Omega}$.
%
%Now, it is clearly monoidal on objects. Moreover, it is monoidal on morphisms since two elements $(e,f)$ and $(e',f')$ of $E\otimes E'$ are equivalent if and only if $e\sim e'$ and $f\sim f'$. Hence $\contract(T\otimes T')=\contract(T)\otimes\contract(T')$.
%\end{proof}
%
%\textcolor{red}{Is $\contract$ full and faithful? It is bijective on objects, full (there's a inclusion from Rel to Trisk) but not faithful. Does it have an adjoint? }

%\begin{proposition}
%The category $\triskellsimp{\cat}$ endowed with the monoidal product induced from the disjoint union is equivalent -- as a monoidal category -- to the category of simple $\object{\Omega}$-weighted directed graphs with disjoint source and target sets of vertices.
%\end{proposition}
%
%\begin{proposition}
%The category $\triskellsimp{\cat}$ endowed with the monoidal product induced from the product is equivalent -- as a monoidal category -- to the category of $\object{\Omega}$-weighted relations.
%\end{proposition}

\section{Fock Functor}\label{sec_fock}

We now define a functor from $\rel{\set}$ to itself which we call the \emph{Fock functor}, by analogy with the \emph{antisymmetric Fock functor}. We recall that this functor is induced by the construction of the \emph{Grassman algebra} or \emph{exterior algebra} \cite[{XIX}, §1, page 733]{lang}. 

%\subsection{The antisymmetric Fock functor}

%Given a vector space $V$, we define the vector space $\Lambda V$ as the direct sum 
%$\oplus_{n\in\naturalN} V^{\otimes n}\diagup\mathcal{I}_{n}$
%where $V^{\otimes n}$ denotes the $n$-fold tensor product of $V$ with itself, and $\mathcal{I}_{n}$ is the ideal generated by the equations $x_{1}\otimes x_{2}\otimes \dots \otimes x_{n}=\epsilon(\sigma)x_{\sigma(1)}\otimes x_{\sigma(2)}\otimes\dots\otimes x_{\sigma(n)}$ where $\sigma$ is a permutation over $\{1,\dots,n\}$ and $\epsilon(\sigma)\in\{-1,1\}$ is its signature.

%In particular, if $V$ has finite dimension $k$, any summand for $n>k$ is isomorphic to the underlying field. Hence the sum can be rewritten as $V^{\otimes n}\diagup\mathcal{I}_{n}$.
%If we write $\mathcal{B}$ a base of the vector space $V$, then the set of finite subsets of $\mathcal{B}$ defines a natural basis of the vector space $\Lambda V$. 

\subsection{The relational Fock functor}

For the definition to make sense in our setting, we will need now to suppose that the monoid $\object{\Omega}$ has a structure of a continuous commutative ring, i.e. we need additive inverses in $\object{\Omega}$. It is important to note that such algebraic structures do exist (for instance, taking the ring of formal power series over a group), though they are not as usual as the well-known semiring $\realposN\cup\{\infty\}$. To consider simpler examples, one can restrict to the subcategory of relations (resp. triskells) between \emph{finite} sets. The continuity requirement is then useless and one could consider simple rings, such as the real numbers.% \note{Does this really exists? We'd like simple example. Let's forget about continuity by simply considering the subcategory where objects are finite sets.} %Consequently, we introduce the notation $\Groth$ as a (combined) notation for the \emph{Grothendieck group}\footnote{Given a monoid $\object{\Omega}$, the group $\Groth$ is defined as the set of pairs $(a,b)\in\object{\Omega}$ together with the addition $(a,b)+(a',b')=(a+a',b+b')$ quotiented by the equivalence relation $ (a,b)\sim(a',b')\Leftrightarrow\exists k, a+b'+k=a'+b+k$. When $\object{\Omega}$ is a semiring, the multiplication can be extended to $\Groth$ by defining $(a,b)(a',b')=(aa'+bb',ab'+ba')$. With this multiplication, $\Groth$ has the structure of a ring. Furthermore, if $\object{\Omega}$ is complete, then $\Groth$ is complete (\note{arrrg, make sure this is true, there might be a problem with the equivalence}).} (resp. \emph{ring}) associated with the monoid (resp. semiring) $\object{\Omega}$. %To define the functor properly, we first introduce some notations.

\begin{notation}
Consider a triskell $\trisk$ in $\alltriskells{A}{B}{\object{\Omega}}$, and choose $\bar{a}=\{a_{1},\dots,a_{k}\}\in\finpow(\object{A})$ and $\bar{b}=\{b_{1},\dots,b_{k}\}\in\finpow(\object{B})$. We define the set $M_{\trisk}[\bar{a},\bar{b}]$ as the set of all pairs $(\sigma,\vec{e})$ where $\sigma$ is a permutation and $\vec{e}$ is a sequence of edges such that $s(e_{i})=a_{i}$ and $t(e_{i})=b_{\sigma(i)}$. Then for all such pairs $(\sigma,\vec{e})$ we define the weight $\omega_{\trisk[T]}[\bar{a},\bar{b}](\vec{e})=\prod_{e\in\vec{e}} w(e)$. We then define $\omega_{\trisk[T]}[\bar{a},\bar{b}]=\sum_{(\sigma,\vec{e})\in M_{\trisk[T]}[\bar{a},\bar{b}]} \epsilon(\sigma)\omega_{\trisk[T]}[\bar{a},\bar{b}](\vec{e})$ where $\epsilon(\sigma)$ is the signature of the permutation $\sigma$.
\end{notation}

\begin{remark}
Notice that $\omega_{\trisk[T]}[\bar{a},\bar{b}]$ is the Leibniz formula for the determinant of the submatrix of $M$ defined as $M[\bar{a},\bar{b}]=(m_{a,b})_{a\in\bar{a}, b\in\bar{b}}$.
\end{remark}

\begin{notation}
For any simple triskell $\trisk[T]$ with source object $\object{A}$ and target object $\object{B}$, we denote by $a\sim_{\omega} b$ the existence of $e\in \object{E}$ with $s(e)=a$, $t(e)=b$ and $w(e)=\omega$. Since $\trisk$ is supposed simple, at most one such $e\in\object{E}$ exists given a pair $(a,b)\in\object{A}\times\object{B}$.
\end{notation}

\begin{definition}
We define the \emph{Fock} functor $\fock$ from the category $\rel{\set}$ to itself as follows (we consider morphisms defined as triskells).
On objects $\fock$ acts as the finite powerset functor: $\object{A}\mapsto\finpow(\object{A})$. On morphisms, we have $\bar{a}=\{a_{1},\dots,a_{k}\}\sim_{w}\{b_{1},\dots,b_{k'}\}=\bar{b}$ in $\fock(T)$ if and only if $k=k'$, $S_{\trisk[T]}[\bar{a},\bar{b}]\neq\emptyset$ and $w=w_{\trisk[T]}[\bar{a},\bar{b}]$.
\end{definition}

\begin{figure*}
\centering
\framebox{
\subfloat[Relation $R$]{
\begin{tikzpicture}[x=0.8cm,y=0.6cm]
	\node (1) at (0,0) {$1$};
	\node (2) at (2,0) {$2$};
	\node (3) at (0,-2) {$4$};
	\node (4) at (2,-2) {$5$};
	\node (fantom) at (0,-2.268) {};

	\draw[-] (1) -- (3) node [midway,right] {\scriptsize{$a$}};
	\draw[-] (1) -- (4) node [near end,above] {\scriptsize{$b$}};
	\draw[-] (2) -- (3) node [near end,above] {\scriptsize{$c$}};
	\draw[-] (2) -- (4) node [midway,left] {\scriptsize{$d$}};
\end{tikzpicture}
}
}
\framebox{
\subfloat[Relation $\fock(R)$]{
\begin{tikzpicture}[x=0.8cm,y=0.6cm]
	\node (0) at (-1,0) {$\emptyset$};
	\node (1) at (0,0) {$\{1\}$};
	\node (2) at (2,0) {$\{2\}$};
	\node (12) at (3,0) {$\{1,2\}$};
	
	\node (00) at (-1,-2) {$\emptyset$};
	\node (3) at (0,-2) {$\{3\}$};
	\node (4) at (2,-2) {$\{4\}$};
	\node (34) at (3,-2) {$\{3,4\}$};

	\draw[-] (0) -- (00) node [midway,right] {\scriptsize{$1$}};
	\draw[-] (1) -- (3) node [midway,right] {\scriptsize{$a$}};
	\draw[-] (1) -- (4) node [near end,above] {\scriptsize{$b$}};
	\draw[-] (2) -- (3) node [near end,above] {\scriptsize{$c$}};
	\draw[-] (2) -- (4) node [midway,left] {\scriptsize{$d$}};
	\draw[-] (12) -- (34) node [midway] {\scriptsize{$ad-bc$}};
\end{tikzpicture}
}
}
\framebox{
\subfloat[Relation $Q$]{
\begin{tikzpicture}[x=0.8cm,y=0.6cm]
	\node (1) at (-2,0) {$1$};
	\node (2) at (0,0) {$2$};
	\node (3) at (2,0) {$3$};
	\node (4) at (-2,-2) {$4$};
	\node (5) at (0,-2) {$5$};
	\node (6) at (2,-2) {$6$};
	\node (fantom) at (0,-2.268) {};

	\draw[-] (1) -- (4) node [midway,right] {\scriptsize{$a$}};
	\draw[-] (1) -- (5) node [near start,above] {\scriptsize{$b$}};
%	\draw[-] (1) -- (6) node [midway,below] {$c$};
	\draw[-] (2) -- (4) node [near start,above] {\scriptsize{$d$}};
	\draw[-] (2) -- (5) node [near end,above,right] {\scriptsize{$e$}};
%	\draw[-] (2) -- (6) node [near start,right] {$f$};
	\draw[-] (3) -- (4) node [near start,above] {\scriptsize{$g$}};
%	\draw[-] (3) -- (5) node [near start,right] {$h$};
	\draw[-] (3) -- (6) node [midway,left] {\scriptsize{$i$}};
\end{tikzpicture}
}
}\\
\framebox{
\subfloat[Relation $\fock(Q)$]{
\begin{tikzpicture}[x=0.9cm,y=0.6cm]
	\node (0) at (-8,0) {$\emptyset$};
	\node (1) at (-7,0) {$\{1\}$};
	\node (2) at (-5,0) {$\{2\}$};
	\node (3) at (-3,0) {$\{3\}$};
	\node (12) at (-2,0) {$\{1,2\}$};
	\node (23) at (0,0) {$\{2,3\}$};
	\node (13) at (2,0) {$\{1,3\}$};
	\node (123) at (3,0) {$\{1,2,3\}$};
	\node (fantom) at (-8.16,0) {};
	\node (fantom2) at (3.66,0) {};
	
	\node (00) at (-8,-2) {$\emptyset$};
	\node (4) at (-7,-2) {$\{4\}$};
	\node (5) at (-5,-2) {$\{5\}$};
	\node (6) at (-3,-2) {$\{6\}$};
	\node (45) at (-2,-2) {$\{4,5\}$};
	\node (56) at (0,-2) {$\{5,6\}$};
	\node (46) at (2,-2) {$\{4,6\}$};
	\node (456) at (3,-2) {$\{4,5,6\}$};
	
	\draw[-] (0) -- (00) node [midway,right] {\scriptsize{$1$}};
	\draw[-] (1) -- (4) node [midway,left] {\scriptsize{$a$}};
	\draw[-] (1) -- (5) node [near start,above] {\scriptsize{$b$}};
%	\draw[-] (1) -- (6) node [midway,below] {$c$};
	\draw[-] (2) -- (4) node [near start,above] {\scriptsize{$d$}};
	\draw[-] (2) -- (5) node [near end,right] {\scriptsize{$e$}};
%	\draw[-] (2) -- (6) node [near start,right] {$f$};
	\draw[-] (3) -- (4) node [near start,above] {\scriptsize{$g$}};
%	\draw[-] (3) -- (5) node [near start,right] {$h$};
	\draw[-] (3) -- (6) node [midway,left] {\scriptsize{$i$}};
	\draw[-] (12) -- (45) node [near start] {\scriptsize{$ae-bd$}};
	\draw[-] (23) -- (45) node [midway,left] {\scriptsize{$-eg$}};
	\draw[-] (13) -- (45) node [near end,below] {\scriptsize{$-bg$}};
%	\draw[-] (12) -- (56) node [midway,above] {$$};
	\draw[-] (23) -- (56) node [near start] {\scriptsize{$bi$}};
	\draw[-] (13) -- (56) node [near end,above] {\scriptsize{$ei$}};
%	\draw[-] (12) -- (46) node [midway,above] {$$};
	\draw[-] (23) -- (46) node [near end,above] {\scriptsize{$ai$}};
	\draw[-] (13) -- (46) node [near start] {\scriptsize{$di$}};
	\draw[-] (123) -- (456) node [midway] {\scriptsize{$aei-bdi$}};

\end{tikzpicture}
}
}
\caption{The functor $\fock$ on two examples}\label{examplefock}
\end{figure*}
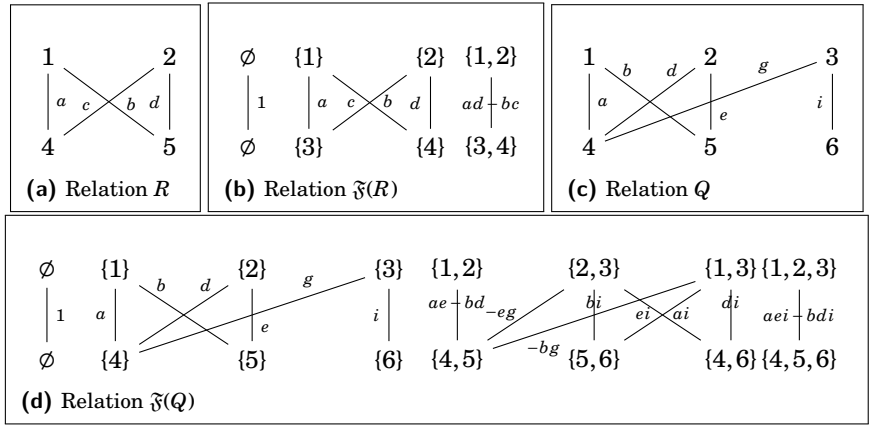

\begin{example}
Let us consider the relation $R$ corresponding to the matrix $\left(\begin{array}{cc} a & b\end{array}\right)$ between $\{1,2\}$ and $\{3\}$. Then $\fock(R)$ is a relation between $\{\emptyset,\{1\},\{2\},\{1,2\}\}$ and $\{\emptyset,\{3\}\}$ defined as follows. First, $\emptyset$ is in relation with $\emptyset$ with weight $1$. Then $\{1\}$ and $\{2\}$ are both in relation with $\{3\}$ with respective weights $a$ and $b$. \autoref{examplefock} illustrates two more involved examples.
\end{example}

%Why call it a Fock functor? Because it is in fact the antysymmetric fock functor, in the sense that the functor we just defined is the image of the antysymmetric fock functor on countable-basis vector spaces, restricted to the bases of vector spaces. i.e.\ if $V$ is a vector space with a basis $\mathcal{B}$, then the Fock functor will map it to the vector space $\Lambda(V)$ which has as natural basis $\textnormal{P}(\mathcal{B})$, the powerset of $\mathcal{B}$. If we consider only the restriction to bases of the action of the antisymmetric Fock functor on linear maps, we obtain that the image of a map $f$ is the powerset map $\textnormal{P}(f)$.

\begin{theorem}\label{fock_monoidal}
The functor $\fock$ is strict monoidal from $(\rel{\set},\coprodmon,\emptyset)$ to $(\rel{\set},\prodmon,1)$.% \textcolor{red}{Careful about weak pullbacks -- but maybe it is alright with the restriction to simple triskells}
\end{theorem}

%\textcolor{red}{Indeed, the image of the pullback $\finpow(A\times_{C} B)$ should only be surjectively mapped to the pullback $\finpow(A)\times_{\finpow(C)}\finpow(B)$. However, maybe the contraction of $\finpow(A\times_{C} B)$ is isomorphic to $\finpow(A)\times_{\finpow(C)}\finpow(B)$; there is something to watch for weights if this is the case.}
%
%\begin{proof}
%Again, we first check that it is indeed a functor.  We only need to verify that the pullback $P$ of $\finpow(f)$ and $\finpow(g)$ is isomorphic to $\finpow(h)$ where $h$ denotes the pullback of $f,g$. The universal property of pullbacks provides a morphism $u$ from $P$ to $\finpow(h)$. We need to show that this morphisms is surjective (the functor only preserves weak pullbacks) [in set, a cone $W$ is a weak pullback if and only if there is a surjective map from $W$ to the pullback].
%\end{proof}

%\section{Products}
%
%This part is not clear yet.
%
%We consider the "free vector space enrichment" of the category $\triskell{\cat}$, i.e.\ morphisms can now be linear combinations of triskells. This defines a construction on morphisms as follows. If $f$ is a morphism in $\homset{A}{B}$ and $g$ is a morphism in $\homset{A'}{B'}$, we can define $f+g$ as $f\otimes 0+0\otimes g$, a morphism in $\homset{A$\otimes$ A'}{B$\otimes$ B'}$. We can also define a morphism in $\homset{A$\otimes$ A}{A}$ as $a,a'\mapsto a+a'$. In other words, this defines a coproduct, compatible with the tensor.

%The image of this coproduct by the Fock functor is the usual coproduct $\oplus$.

This shows that the image of $(\triskell{\set},\oplus,0)$ through the composition $\fock\circ\contract$ is the category $(\rel{\set},\otimes,1)$. I.e. the Interaction Graphs' interpretation $\Int{\pi}{{\textsc{ig}}}$ of a \MLL proof $\pi$ is mapped to its weigthed relational model's interpretation $\Int{\pi}{{\textsc{wr}}}$. This result, however, can be improved by lifting the Fock functor to the category of triskells. 

\subsection{The Triskells Fock Functor}

We will now show how to lift the fock functor from simple triskells to general triskells, i.e. we will show there exists a functor $\liftedfock$ such that the following diagram commutes. We will define $\triskell{\cat}^{0}$ later on; let us point out for now that it is a quotient of $\triskell{\cat}$ w.r.t. a simple congruence.
\vspace{-0.5cm}
\begin{center}
\begin{tikzpicture}[x=1cm,y=0.6cm]
	\node (TD) at (0,2) {$(\triskell{\cat},\disjun)$};
	\node (TP) at (4,2) {$(\triskell{\cat},\times)^{0}$};
	
	\node (RD) at (0,0) {$(\rel{\cat},\disjun)$};
	\node (RP) at (4,0) {$(\rel{\cat},\times)$};
	
	\draw[->] (TD) -- (RD) node [midway,left] {$\contract$};
	\draw[->] (TP) -- (RP) node [midway,left] {$\contract$};
	\draw[->] (RD) -- (RP) node [midway,above] {$\fock$};
	\draw[dotted,->] (TD) -- (TP) node [midway,above] {$\liftedfock$};
\end{tikzpicture}
\end{center}

Although it is necessary that $\object{\Omega}$ have the structure of a complete ring (or simply a ring if we restrict to the subcategories where objects are finite sets) for the above diagram to commute, we will in fact not need much more than a monoid structure on $\object{\Omega}$ for defining $\liftedfock$. This is yet another argument in favour of considering triskells instead of relations. This lifted functor is defined in a way which is similar to the definition $\fock$, but simpler as one does not perform any summations. I.e. every element of $M_{\trisk}[\bar{a},\bar{b}]$ will give rise to a new edge. The only thing we will need are signs, i.e. ways to interpret $\epsilon(\sigma)$ for a permutation $\sigma$. This is done by picking a monoid $\object{\Omega}$ that can be written as the product of a monoid with the monoid $\{-1,1\}$ (with multiplication). In the following, we will call such structure a \emph{signed monoid}.

\begin{remark}
We chose here to fix $\object{\Omega}$ as a signed monoid, but we could have as well defined $\liftedfock$ as a functor from $\triskell{\set}$ to $\triskell[\{-1,1\}\times\object{\Omega}]{\set}$ for any monoid $\object{\Omega}$.
\end{remark}

%\note{$S_{\trisk}[\bar{a},\bar{b}]$ becomes a multiset. Adapt.}

\begin{definition}
We define $\liftedfock$ as follows. On objects, $\liftedfock$ acts as the finite powerset functor. On morphisms, it maps the triskell $T$ to the triskell $\liftedfock(T)$ defined as follows: for each $(\sigma,\vec{e})$ in $M_{\trisk[T]}[\bar{a},\bar{b}]$, there is exactly one edge of source $\bar{a}$, target $\bar{b}$ and weight $\epsilon(\sigma)\omega_{\sigma}[\bar{a},\bar{b}](\vec{e})$.
%There are edges from $\{a_{1},a_{2},\dots,a_{l}\}$ to $\{b_{1},b_{2},\dots,b_{k}\}$ if and only if $k=l$. Moreover, there is one such edge for each set of edges $e_{1},\dots,e_{k}$ such that there is a permutation $\sigma$ of $\{1,\dots,k\}$ with $s(e_{i})=a_{i}$ and $t(e_{i})=b_{\sigma(i)}$. The weight of this edge is then computed as the product of the weights $w(e_{i})$ multiplied by $\epsilon(\sigma)$, the signature of the permutation.
\end{definition}

%\note{This defines a monoidal functor turning the disjoint union into a product. What happens with the trace then? The domain category is traced monoidal; the codomain category is also traced monoidal. The image of the trace is however not a priori the same as the trace on the codomain category. Is it another trace? What becomes of these in the contraction?}

It turns out that $\liftedfock$ is \emph{not} a functor from $\triskell{\cat}$ to itself. To understand this, and understand why the same situations are not a problem with relations, let us consider the triskells (actually these are relations) $\trisk$ and $\trisk'$ shown in \autoref{triskellA} and \autoref{triskellB}. Then the fact that $\liftedfock$ is not a functor can be seen by computing $\liftedfock(\trisk'\circ\trisk)$ (\autoref{Fig3A}) and $\liftedfock(\trisk')\circ\liftedfock(\trisk)$ (\autoref{Fig3B}). As the reader can see for herself, the contraction operation will cancel the edges from $\{1,2\}$ to $\{4,5\}$ in the triskell $\liftedfock(\trisk'\circ\trisk)$. This is why $\fock$ is indeed a functor while $\liftedfock$ is not.

\begin{figure*}
\centering
\framebox{
\subfloat[Triskell $\trisk$ \label{triskellA}]{
\begin{tikzpicture}[x=0.7cm,y=0.6cm]
	\node (1) at (0,0) {$1$\textcolor{white}{\}}};
	\node (2) at (2,0) {$2$\textcolor{white}{\}}};
	\node (3) at (1,-2) {$3$\textcolor{white}{\}}};

%	\node (fantom) at (0,-2.268) {};

	\draw[-] (1) -- (3) node [midway,left] {\scriptsize{$a$}};
	\draw[-] (2) -- (3) node [midway,right] {\scriptsize{$b$}};
\end{tikzpicture}
}
}
\framebox{
\subfloat[Triskell $\trisk'$\label{triskellB}]{
\begin{tikzpicture}[x=0.7cm,y=0.6cm]
	\node (3) at (1,-2) {$3$\textcolor{white}{\}}};
	\node (4) at (0,-4) {$4$\textcolor{white}{\}}};
	\node (5) at (2,-4) {$5$\textcolor{white}{\}}};

	\draw[-] (3) -- (4) node [midway,left] {\scriptsize{$c$}};
	\draw[-] (3) -- (5) node [midway,right] {\scriptsize{$d$}};
\end{tikzpicture}
}
}
\framebox{
\subfloat[Triskell $\liftedfock(\trisk'\circ\trisk)$ \label{Fig3A}]{
\begin{tikzpicture}[x=0.7cm,y=0.6cm]
	\node (0) at (-1,0) {$\emptyset$};
	\node (1) at (0,0) {$\{1\}$};
	\node (2) at (2,0) {$\{2\}$};
	\node (12) at (3,0) {$\{1,2\}$};
	
	\node (00) at (-1,-2) {$\emptyset$};
	\node (3) at (0,-2) {$\{4\}$};
	\node (4) at (2,-2) {$\{5\}$};
	\node (34) at (3,-2) {$\{4,5\}$};

	\draw[-] (0) -- (00) node [midway,right] {\scriptsize{$1$}};
	\draw[-] (1) -- (3) node [midway,right] {\scriptsize{$a$}};
	\draw[-] (1) -- (4) node [near end,above] {\scriptsize{$b$}};
	\draw[-] (2) -- (3) node [near end,above] {\scriptsize{$c$}};
	\draw[-] (2) -- (4) node [midway,left] {\scriptsize{$d$}};
	\draw[-] (12) -- (34) node [near end] {\scriptsize{$abcd$}};
	\draw[-] (12) .. controls (4,0) and (4,-2) .. (34) node [near start] {\scriptsize{$-abcd$}};
\end{tikzpicture}
}
}
\framebox{
\subfloat[Triskell $\liftedfock(\trisk')\circ\liftedfock(\trisk)$ \label{Fig3B}]{
\begin{tikzpicture}[x=0.7cm,y=0.6cm]
	\node (0) at (-1,0) {$\emptyset$};
	\node (1) at (0,0) {$\{1\}$};
	\node (2) at (2,0) {$\{2\}$};
	\node (12) at (3,0) {$\{1,2\}$};
	
	\node (00) at (-1,-2) {$\emptyset$};
	\node (3) at (0,-2) {$\{4\}$};
	\node (4) at (2,-2) {$\{5\}$};
	\node (34) at (3,-2) {$\{4,5\}$};

	\draw[-] (0) -- (00) node [midway,right] {\scriptsize{$1$}};
	\draw[-] (1) -- (3) node [midway,right] {\scriptsize{$a$}};
	\draw[-] (1) -- (4) node [near end,above] {\scriptsize{$b$}};
	\draw[-] (2) -- (3) node [near end,above] {\scriptsize{$c$}};
	\draw[-] (2) -- (4) node [midway,left] {\scriptsize{$d$}};
%	\draw[-] (12) -- (34) node [midway] {\scriptsize{$ad-bc$}};
\end{tikzpicture}
}
}
\caption{$\liftedfock$ is not a functor\label{notafunctor}}
\end{figure*}

However, one can show that $\liftedfock$ is a functor \emph{up to some equivalence} on morphisms. For this, we define a congruence on the set of triskells over $\{1,-1\}\times\object{\Omega}$ as follows: two triskells $\trisk$ and $\trisk'$ are equivalent if and only if there exists two \enquote{zero triskells} $\trisk[Z]$ and $\trisk[Z']$ and a triskell $\trisk[A]$ such that $\trisk=\trisk[A]\cup\trisk[Z]$ and $\trisk'=\trisk[A]\cup\trisk[Z']$. Here a \emph{zero triskell} is a triskell whose set of edges $E$ can be decomposed as $E^{+}+E^{-}$ so that there is a bijection $\theta:E^{+}\rightarrow E^{-}$ satisfying $s(e)=s(\theta(e))$, $t(e)=t(\theta(e))$ and $w(e)=-w(\theta(e))$. Then $\liftedfock$ can be shown to be a functor from $\triskell{\set}$ to the quotient $\triskell{\set}^{0}$ of $\triskell{\set}$ by this congruence, and this congruence is the smallest that makes $\liftedfock$ a functor. Notice moreover that this congruence is contained into the equivalence defined from the contracting functor $\contract$, i.e. the equivalence defined by $\trisk\sim\trisk'$ if and only if $\contract(\trisk)=\contract(\trisk')$. This implies that the functor $\contract$ is well-defined on the quotient category $\triskell{\set}^{0}$, as two equivalent triskells in $\triskell{\set}$ have the same image through $\contract$. Lastly, one easily checks that this quotient is compatible with the monoidal structure, i.e. $(\triskell{\set}^{0},\prodmon,1)$ is a monoidal category.

\begin{theorem}\label{fock_trik_monoidal}
The functor $\liftedfock$ is strict monoidal from $(\triskell{\set},\coprodmon,\emptyset)$ to $(\triskell{\set}^{0},\prodmon,1)$, and $\fock\circ\contract=\contract\circ\liftedfock$.% \textcolor{red}{Careful about weak pullbacks -- but maybe it is alright with the restriction to simple triskells}
\end{theorem}

A consequence of \autoref{fock_monoidal} is the fact that a \MLL proof's interpretation in the dynamic models is mapped, through $\fock\circ\contract$, to the interpretation of the same proof in the relational model, modulo the fact that we are working with a continuous semiring (or a ring when considering only finite sets as objects). More importantly, this last theorem \ref{fock_trik_monoidal} extends this result when one considers triskells instead relations. I.e. a \MLL proof's interpretation in the interaction graphs model for a monoid $\object{\Omega}$ is mapped through $\liftedfock$ to the interpretation of the same proof in the triskells denotational model for the monoid $\{-1,1\}\times\object{\Omega}$ (modulo the simple quotient considered above). We will now show how this result actually extends to the double-glueing constructions.

\section{Orthogonalities}\label{sec_orthogonality}

\subsection{Some preliminary remarks}

In this section, we start discussing the image of the double-glueing construction considered in Interaction Graphs. First, we limit our discussion to the subcategory of $\rel{\set[\aleph_{1}]}$ where all objects are finite sets. A further restriction will be that $\object{\Omega}$ is chosen as a subring of the complex numbers $\complexN$. These constraints are quite strong but they were introduced by the author \cite{seiller-goim} to obtain a  combinatorial version of Girard's hyperfinite geometry of interaction \cite{goi5}. 

In this particular case, we can define in the category of relations both the (normalised) trace -- i.e.\ the mean of diagonal coefficients -- and the determinant of matrices. 
%\begin{itemize}
%\item the normalised trace of a morphism as a natural generalisation of the trace of matrices. i.e.\ for $M\in\homset{A}{A}$, $\tr(M)$ is equal to the mean of diagonal coefficients, i.e.\ $K\times \tr(m)=\sum_{i\in \object{A}} m_{i,i}$ where $K$ is the cardinality of $A$;
%\item the determinant of a morphism, either directly, or using the known equality $\log(\det(M))=\tr(\log(M))$. 
%\end{itemize}
We now recall the following well-known theorem from linear algebra.

\begin{theorem}\label{thm_matdettrace}
Let $A$ be a matrix with complex coefficients, then $ \tr(\fock(A))=\det(1+A)$.
\end{theorem}

%\begin{proof}
%As a complex matrix of dimension $k\times k$, $A$ can be diagonalized. It is then an easy check: the determinant of $1+A$ is the product of eigenvalues $\prod_{i=1}^{k} (1+\lambda_{i})$ which can be developed as $\sum_{P\subset \{1,\dots,k\}}\sum_{i\in P} \lambda_{i}$, i.e.\ the trace of $\fock(A)$.
%\end{proof}

%\begin{corollary}
%$$ \tr(\fock(A)\fock(B))=-\det(1-AB) $$
%\end{corollary}
%
%\begin{proof}
%$$ -\tr(\fock(A)\fock(B))=-\tr(\fock(AB)))=\tr(\fock(-AB))=\det(1+(-AB))=\det(1-AB) $$
%\end{proof}

%As a consequence, $\det(1-AB)$ belongs to $]0,1[$ if and only if  the trace of $\fock(A)\fock(-B)$ belongs to $]0,1[$.

We now fix the subring $[0,1]$ of the complex numbers as our weight object $\object{\Omega}$, and we restrict to matrices with operator norm at most $1$. This second restriction implies that the operations interpreting multiplicative linear logic will not create weights greater than $1$ \cite{seiller-goim}, i.e. the fact that $[0,1]$ is not a subring (i.e. not closed under addition) will not induce incoherences in the model.\footnote{The norm restriction allows to bypass the ring structure as it insures that the operations interpreting linear logic connectives will not create triskells (or graphs) with weights outside $[0,1]$. This is clear for the disjoint union used to interpret the tensor product, it is proved for the trace in earlier works \cite{seiller-goim}.} These restrictions still provide an adequate framework for defining the combinatorial variant of Girard's hyperfinite model \cite{goi5,seiller-goim} and probabilistic coherent spaces.

Girard's hyperfinite GoI uses the Fuglede-Kadison determinant $\det^{\mathrm{\scriptsize{FK}}}$ \cite{FKdet}, a (positive-)real-valued determinant which corresponds, in finite dimensions, to the absolute value of the usual determinant (to the power $1/k$). This determinant is used to define the orthogonality considered in Girard's GoI models, i.e. two matrices $A,B$ are (det-)orthogonal, noted $\goiorth$, when $\det^{\mathrm{\scriptsize{FK}}}(1-AB)\neq 0,1$, i.e.\ when $\abs{\det(1-AB)}\neq 0,1$. The trace, on the other hand, can be used to define (a variant of) the orthogonality considered in probabilistic/quantum coherence spaces, i.e. two matrices $A,B$ are (trace-)orthogonal, denoted $\cohorth$, when $\tr(AB)\in]0,1[$.\footnote{The original definition allows for the values $0$ and $1$ which are here forbidden. This slight change however do not impact the result, i.e. we still define from this orthogonality a model of multiplicative linear logic.}

The theorem above shows that $\det^{\mathrm{\scriptsize{FK}}}(1-AB)=\abs{\tr(\fock(AB))}$, i.e. when we restrict to norm 1 matrices with coefficients in $[0,1]$:
$$ 
A\goiorth B  				~\Longleftrightarrow~	\det{}^{\mathrm{\scriptsize{FK}}}(1-AB)\in]0,1[
							~\Longleftrightarrow~	\abs{\tr(\fock(A)\fock(B))}\in]0,1[
							~\Longleftrightarrow~    \fock(A)\cohorth\fock(B)	
$$

This implies that the interaction graph model detailed in the author's first work \cite{seiller-goim} is mapped to the model of \emph{strict} coherent spaces through the functor $\fock$. This result, while interesting on its own, can be generalised if one works with triskells instead of relations. The reason triskells provide a better framework is that we will be able to get rid of the restrictions we imposed in this section. Indeed, we asked for objects to be restricted to finite sets: the reason for this is that dealing with infinite sets implies dealing with infinite sums. Thus infinite sets force us out of the usual domain of definition of matrix determinants. The second restriction, i.e. that the weight object is a subring of the complex numbers, is also linked to the definition of the determinant. Indeed, the definition of a determinant for matrices over arbitrary semirings is quite involved and has no clear solution \cite{detsemiring}. Lastly, the condition on the norm comes from the very definition of Girard's model: execution cannot be defined in general when the norm is greater than $1$ \cite{seiller-masas}.

We will now show how one can define a much more general and satisfying correspondence when dealing with triskells. Such a generalisation will provide many more models since there are fewer restrictions on the weight object $\object{\Omega}$. This is done by introducing notions of determinant and trace of a triskell that generalise the usual definition for complex matrices. We ask but one thing about these extensions, namely that they map the determinant to the trace through $\liftedfock$. That is, we ensure that a generalisation of \autoref{thm_matdettrace} holds. %Moreover, we will verify that the definition of determinant thus generalised coincides with the measurement on graphs considered in interaction graphs \cite{seiller-goiadd}. This will allow for the definition of coherence space semantics over the category of triskells that are the image of the author's interaction graphs models.

\subsection{Determinant and trace for triskells}

The remark above motivates a definition of determinant and trace in the categories of triskells. This can be done in a coherent way, namely such that the equivalent of \autoref{thm_matdettrace} holds. %Note also that the determinant defined below corresponds to the measurement considered in the authors' earlier works \cite{seiller-goiadd}. For instance, if $\meas{\cdot,\cdot}$ denotes this measurement, we have $\meas{A,B}=\det{}_{m}(1-AB)$, modulo the identification of graphs and triskells. 
Note also that the trace satisfies the property that $\tr_{m}(AB)=\tr_{m}(BA)$, and will be used to define a notion of quantitative coherent spaces in $\triskell{\set}$.

Let $(\object{K},+)$ be a complete monoid -- i.e. a monoid with a notion of infinite sums --, let $\object{\Omega}$ be a signed monoid, and $m:\object{\Omega}\rightarrow\object{K}$ be any map, i.e.\ not necessarily a monoid (or equivalently group) homomorphism. We define the $m$-trace and the $m$-determinant of any triskell with same source and target objects the finite set $A$ as
$$\tr_{m}(\trisk[T])=\sum_{e\in E, s(e)=t(e)} m(w(e))
\hspace{2cm}
\det{}_{m}(\trisk[T])=\sum_{\sigma\in\mathfrak{G}_{A}} \sum_{(e_{i})\in \prod_{i\in A} E[\sigma(i),i]} m(\epsilon(\sigma)\prod w(e_{i}))$$
In the last expression, $\mathfrak{G}_{A}$ should be understood as the set of permutation over $A$, and for all $x,y\in A$ we denote by $E[x,y]$ the subset of edges $e$ such that $s(e)=x$ and $t(e)=y$.

\begin{remark}
If one keeps to the subcategory where objects are finite sets, then completeness of $\object{K}$ is not necessary and any monoid would do. As a particular case, the identity map on complex numbers allows one to recover the usual determinant and trace of matrices.
\end{remark}

%If $K$ is a complete semiring, teh expression this last expression makes sense only when $A$ is finite, since when $A$ is infinite we find ourselves with infinite products to deal with. Before proving the theorem, let us discuss the extension to triskells over infinite sets.  

Although the trace defined as it is above can be used to deal with matrices over infinite sets, the determinant considered there cannot, as one needs to define infinite products for the expression to make sense. Given a monoid $\Omega$, we define the extended monoid $\Omega^{0}$ by adding an absorbing element $0$, i.e. $0.a=a.0=0$ for all $a\in\Omega^{0}$. One can then define, in this extended monoid, a satisfying notion of infinite products, namely $\prod_{i\geqslant 0}a_{i}$ is defined equal to $\prod_{i=0}^{k} a_{i}$ if for all $j>k$ we have $a_{j}=1$ and equal to $0$ otherwise. By implicitely considering elements of $\Omega$ as their image in $\Omega^{0}$ and extending $m$ to $\Omega^{0}$ by defining $m(0)$ as being equal to the unit of the sum in $K$, the expression above for determinant makes sense for arbitrary triskells. With this definition of determinant and trace, we obtain the following theorem.

%\note{This could be dealt with by considering as $\object{K}$ a semiring with infinite product (for more details we refer to so-called complete semiring semimodule pairs \cite{completesemiringsemimodulepairs}). For instance, one could pick $K$ as the complete semiring $\realposN\cup\{\infty\}$ with the infinite products $\prod_{i\geqslant 0}a_{i}$ defined as equal to $0$ if there exists $i$ with $a_{i}=0$, equal to $\prod_{i=0}^{k} a_{i}$ if for all $j>k$ we have $a_{j}=1$, and equal to $\infty$ otherwise.}

%We propose here another alternative, namely extending $\object{\Omega}$ and $\object{K}$ with absorbing elements written $0$. Then one defines an infinite product $\prod_{i\geqslant 0}a_{i}$ in $\object{\Omega^{0}}$ as equal to $\prod_{i=0}^{k} a_{i}$ if for all $j>k$ we have $a_{j}=1$ and equal to $0$ otherwise. Moreover, we consider that $m:\object{\Omega}\rightarrow\object{K}$ preserves absorbing elements (\note{do we need to ask for that? yes!})
%$$\det{}_{m}(\trisk[T])=\sum_{\sigma\in\mathfrak{G}^{\textnormal{fin}}_{A}} \sum_{K\in \finpow{A}, K\supset \textnormal{supp}(\sigma)}\sum_{(e_{i})\in \prod_{i\in K} E[\sigma(i),i]} m(\epsilon(\sigma)\prod w(e_{i}))$$

%These definitions allow for the definition of trace and determinants of triskells even when the source and target object is not a finite set. For this, one keeps exactly the same definition of trace, and sums over finite permutations 

\begin{theorem}\label{thm_triskdettrace}
For any triskell $\trisk[A]$ with same source and target sets, $\det{}_{m}(1+\trisk[A])=\tr_{m}(\liftedfock(\trisk[A]))$.
\end{theorem}

%\begin{proof}
%There exists a direct proof of this identity in the matrix case based on the Leibniz formula for the determinant. The main technical content of the proof corresponds to showing that $\det(1+A)=\sum_{K\subset\{1,\dots,n\}} \det(A_{K})$ for any $n\times n$ matrix $A$. A simple adaptation of the latter provides a proof of the statement.
%\end{proof}

%\note{A more general version of this theorem can be obtained when considering constraints on $\trisk[A]$. For instance, if $\trisk[A]$ is such that all infinite paths in $\trisk[A]$ go through a finite number of \emph{vertices}. Another, more satisfying, way to extend this theorem to general triskells would be to consider a modified Fock functor where the underlying operation on objects consists in taking the \enquote{full} powerset functor instead of the finite powerset functor, and defined on morphisms in a similar way as the current Fock functor. Doing so however exceeds what can be exposed in this paper because it would require a careful and involved treatment of infinite products for the mere definition of the modified Fock functor. }

%\note{All these results together yield. Theorem. If $K=\realposN$ and $m$ is a monoid morphism, then $\meas{A,B}=-\log(\det(1-A))$.}

\section{Quantitative coherence spaces}

\subsection{The models}

We will describe here a notion of \emph{quantitative coherence spaces}. Although coherence spaces and their probabilistic and quantum variations have been defined as a double-glueing construction on the category of (weighted) relations, we here describe them as a double-glueing construction applied to the category of triskells. This change in the underlying category allows for a wider array of definable models, as the category of weighted relations uses a complete semiring to represent quantitative features, while triskells need only the simpler structure of a monoid. We stress that this double-glueing on triskells gives rise to a double-glueing construction on weighted relations through the contraction functor $\contract$.

\begin{definition}
Given a complete monoid $(\object{K},+)$, a map $m: \object{\Omega}\rightarrow \object{K}$ and a subset $\Bot$ of $\object{K}$, we define a binary relation -- the orthogonality -- on triskells by $\trisk \poll{}{} \trisk[T']\textrm{ if and only if }\tr_{m}(\trisk \circ \trisk[T'])\in\Bot$.
We define the orthogonal of a set of triskells $A$ as the set $A^{\pol}=\{\trisk~|~\forall\trisk[A]\in A, \trisk\poll{}{}\trisk[A]\}$.
\end{definition}

\begin{definition}
A quantitative coherence space (\qcs) $A$ is a pair $(X^{A},\cond{A})$ of a set $X^{A}$ and a  bi-orthogonally closed set $\cond{A}$ of triskells over $X^{A}$ (i.e. of source and target $X^{A}$), i.e. $\cond{A}=(\cond{A}^{\pol})^{\pol}$.
\end{definition}

\begin{definition}
Let $A=(X^{A},\cond{A})$ and $B=(X^{B},\cond{B})$ be \qcs. Their tensor product is defined as
$$A\otimes B=(X^{A}\times X^{B},\{\trisk\times\trisk[T'], \trisk\in\cond{A}, \trisk[T']\in\cond{B}\}^{\pol\pol})$$
\end{definition}

As it is usual with this way of representing coherence spaces, no nice representation of the dual connective $\parr$ can be obtained. The linear implication, however, is nicely characterised. 

\begin{definition}
Let $\trisk[F]$ be a triskell over $X\times Y$, and $\trisk[A]$ be a triskell over $X$. We define the application $\trisk[{[F]A}]$ as the triskell $(E,Y,Y,\Omega,s,t,\omega)$, where:
$$E=\{(f,a)~|~x,x'\in X, y,y'\in Y, f\in E^{F}[(x,y),(x',y')], a\in E^{A}[x,x']\}$$
$$s(f,a)=\pi_{Y}(s(f)) ~~~~~ t(f,a)=\pi_{Y}(t(f)) ~~~~~ \omega(f,a)=\omega(f)\omega(a)$$
\end{definition}

\begin{definition}
If $A=(X^{A},\cond{A})$ and $B=(X^{B},\cond{B})$ are \qcs, the linear arrow is defined as follows:
$$A\multimap B=(X^{A}\times X^{B},\{\trisk[F]\in\alltriskellssym{A\times B}{\Omega}~|~ \forall \trisk[A]\in\cond{A}, \trisk[{[F]A}]\in\cond{B}\})$$
\end{definition}

\begin{theorem}\label{modelMLL}
For any $m$ and $\Bot$, this defines a model of \MLL.
\end{theorem}

\subsection{Relations to previous models}

These models generalise previously introduced coherence spaces models. To see this, we will need some additional definitions. Say a triskell $\trisk=(\object{E},\object{S},\object{T},\object{\Omega},s,t,w)$ is \emph{diagonal} when $\object{S}=\object{T}$ and for all $e\in\object{E}$, $s(e)=t(e)$. Restricting morphisms to diagonal (resp. hermitian) triskells defines a subcategory of the category of triskells. Restricting the definitions of \qcs just considered to this subcategory leads to the usual notions of coherence spaces and probabilistic coherence spaces (in the style of Girard \cite{qcs}). Restricting in the same manner to the straightforward notion of \emph{hermitian} triskells\footnote{A triskell $\trisk=(\object{E},\object{S},\object{T},\object{\Omega},s,t,w)$ is hermitian if $\object{\Omega}=\complexN$, $\object{E}=\object{F}\disjun\object{G}$, and $\exists\phi:\object{F}\rightarrow \object{G}$ bijective s.t. $w(\phi(e))=\bar{w(e)}$.} leads to Girard's quantum coherence spaces model.

\begin{proposition}\label{GirardQCOH}
The restriction of \qcs to \emph{simple diagonal triskells} (and finite objects) yields usual coherence spaces -- when $\Omega=\{1\}$ -- and probabilistic coherence spaces -- when $\Omega=\realposN$ -- (for \MLL). Restriction to \emph{simple hermitian triskells} (and finite objects) leads to quantum coherence spaces (for \MLL).
\end{proposition}

Now, probabilistic coherence spaces were generalised to deal with infinite objects by Danos and Ehrhard; this allowed them to define exponentials, something unavailable in Girard's construction. It turns out that their generalisation is also a special case of our construction. To formalise this, we consider the subcategory of diagonal triskells and the \qcs one can define in this category. To simplify things, first notice that it is not necessary to consider $\Omega=\realposN$: since we are allowing multiple edges with same source and target, all $\realposN$-weighted (simple) triskells can  be obtained from $[0,1]$-weighted triskells through the collapse functor. Secondly, Danos and Ehrhard require two additional conditions that can be translated in our setting by considering the following definition. We will use here \emph{(weighted) partial identities} to express it, i.e. if $B\subset X^{A}$, the partial identity on $B$ weighted by $\lambda\in\object{\Omega}$ is the triskell $(B,X^{A},X^{A},\object{\Omega},\iota,\iota,\lambda\umap[B])$ where $\iota$ denotes the inclusion of $B$ in $X^{A}$ and $\lambda\umap[B]$ is the constant function equal to $\lambda$ (definable through the \enquote{unit} map $\umap[B]$).

\begin{definition}
We say that a \qcs $A$ is \emph{everywhere bounded below} if for every $x\in X^{A}$ there is a $\lambda>0$ such that the partial identity of $x$ weighted by $\lambda$ lies in $\cond{A}$. If both $A$ and $A^{\pol}$ are everywhere bounded below, we say that $A$ is \emph{bounded}.
\end{definition}

\begin{remark}
This restriction to specific bi-orthogonally closed sets is reminiscent of a similar restriction in the \IG models \cite{seiller-goiadd}. It is not clear if the two restrictions are related, but they are strikingly similar at a first glance. A more detailed study of these aspects will have to be relegated to later work, as it is intimately related to the interpretation of additives.
\end{remark}

The constructions for multiplicative connectives in Danos and Ehrhard paper are then the same as the one considered here. The essential point is that the tensor product of bounded \qcs is again a bounded coherence space. This leads to the following proposition.

\begin{proposition}\label{DEQCOH}
The restriction to \emph{bounded \qcs} in the subcategory of \emph{diagonal triskells} (when $\object{\Omega}=[0,1]$) yields Danos and Ehrhard probabilistic coherence spaces model (for \MLL).
\end{proposition}

We have shown how \qcs can be related to previous coherence space semantics. We will now explain how it relates to the author's Interaction Graphs (\IG) models. We first prove a number of properties for the introduced notions of determinant and trace, which will then lead us to \autoref{quantfunctormodel} stating a formal link between the two families of models.

\begin{proposition}\label{lineartrace}
If $(K,\cdot,+)$ is a semiring in which infinite sums are invariant w.r.t. every permutation of the index set (not only finite ones) and $m$ is a monoid morphism from $\object{\Omega}$ to $(\object{K},\cdot)$, i.e. $m(a.b)=m(a).m(b)$, then $\tr_{m}$ is linear, i.e. $\tr(T\disjun T')=\tr(T)+\tr(T')$ and $\tr(a.T)=m(a)\tr(T)$.
\end{proposition}

We can now relate these notions of trace and determinant to the measurement introduced in earlier work on interaction graphs \cite{seiller-goiadd}, and denoted by $\meas{\cdot,\cdot}$. We consider here triskells over finite sets, and the real numbers as the semiring $\object{K}$. Notice the latter is not complete, but completeness is needed only for dealing with triskells over infinite sets.

\begin{proposition}\label{measurementdet}
Suppose $\object{K}=\realN$ and that $m$ is a monoid morphism such that $m(-1)=-1$. Suppose moreover that there is a family $a_{k}\in\object{\Omega}, k\in\naturalN$ with $m(a_{k})=1/k$ and that $A,B$ are triskells over finite sets. Writing $\hat{m}(x)=-\log(1-m(x))$, we have\footnote{We identify in this statement triskells and weighted directed graphs (the measurement being defined \cite{seiller-goiadd} on this type of graphs).}:
$$-\log(\det{}_{m}(1-AB))=\tr_{m}(\sum a_{k}(AB)^{k})=\meas[\hat{m}]{A,B}$$
\end{proposition}

In other words, the orthogonality defined in interaction graphs by $A\poll{}{} B\mathrm{~iff~}\meas[\hat{m}]{A,B}\in\Bot$ and the orthogonality of \qcs are mapped one to the other when we restrict to the hypotheses of the theorem. This yields the following theorem.

\begin{theorem}\label{quantfunctormodel}
The lifted Fock functor maps the \IG model (for \MLL) to \qcs models (for \MLL) defined on its image. 
\end{theorem}

\subsection{Beyond multiplicatives}

It is natural to wonder about the generalisation of \qcs to larger fragments of linear logic. First, the model extends quite straightforwardly to additives. 

\begin{definition}
Given \qcs $A=(X^{A},\cond{A})$ and $B=(X^{B},\cond{B})$, we define
$$ A\with B=(X^{A}\disjun X^{B},\{\trisk[A+B]~|~\trisk[A]\in\cond{A}, \trisk[B]\in\cond{B}\}^{\pol\pol}) $$
\end{definition}

As usual, the dual $(A^{\pol}\with B^{\pol})^{\pol}$ is denoted $A\oplus B$. With this definition, \autoref{GirardQCOH} and \autoref{DEQCOH} are easily extended to the \MALL fragment. However, the most interesting question concerns the extension to exponential connectives. It turns out that one can define exponentials using the \emph{symmetric Fock functor}.

\begin{definition}
We define $\symmfunctor$ as follows. On objects, $\symmfunctor$ acts as the finite multisets $\mfin(\cdot)$ functor. On morphisms, the functor maps the triskell $T$ to the triskell $\symmfunctor(T)$: for each $(\sigma,\vec{e})$ in $M_{\trisk[T]}[\bar{a},\bar{b}]$, there is exactly one edge in $\symmfunctor(T)$ of source $\bar{a}$, target $\bar{b}$ and weight $\omega_{\sigma}[\bar{a},\bar{b}](\vec{e})$.
%There are edges from $\{a_{1},a_{2},\dots,a_{l}\}$ to $\{b_{1},b_{2},\dots,b_{k}\}$ if and only if $k=l$. Moreover, there is one such edge for each set of edges $e_{1},\dots,e_{k}$ such that there is a permutation $\sigma$ of $\{1,\dots,k\}$ with $s(e_{i})=a_{i}$ and $t(e_{i})=b_{\sigma(i)}$. The weight of this edge is then computed as the product of the weights $w(e_{i})$ multiplied by $\epsilon(\sigma)$, the signature of the permutation.
\end{definition}

\begin{proposition}\label{symmmonoidal}
The functor $\symmfunctor$ is strict monoidal from $(\triskell{\set},\coprodmon,\emptyset)$ to $(\triskell{\set},\prodmon,1)$.
\end{proposition}

\begin{definition}
Given a \qcs $A=(X^{A},\cond{A})$ we define the following \qcs.
$$ \oc A = (\mfin(X^{A}),\{\symmfunctor(\trisk[A])~|~\trisk[A]\in\cond{A}\}^{\pol\pol}) $$
%As usual, $(\oc A^{\pol})^{\pol}$ will be denoted $\wn A$.
\end{definition}

\begin{theorem}\label{modelLL}
For any $m$ and $\Bot$, this defines a model of \LL.
\end{theorem}

Let us notice that this definition of exponentials is similar to that of exponentials in weighted relations models of Laird \emph{et al.} \cite{quantdenot}. More formally, given a triskell $\trisk$ contracting to the weighted relation $R_{\trisk}$, the triskell obtained by applying the symmetric Fock functor $\symmfunctor \trisk$ is mapped, through the contraction functor, to the relation denoted by $\oc R_{\trisk}$ in Laird \emph{et al.}

We finally explain how this differs from the exponentials considered by Danos and Ehrhard \cite{probcoh}. First, we remark that if $\trisk$ is a diagonal triskell, then $\symmfunctor\trisk$ will be diagonal as well. Then starting from a simple diagonal triskell $\trisk$ corresponding to the relation $R$ over a set $\object{X}$ with weights in $[0,1]$, we can compute the triskell $\contract\symmfunctor \trisk=(w_{m})_{m\in\mfin(\object{X})}$ and compare it to Danos and Ehrhard exponential $\oc R=(R^{!}_{m})_{m\in\mfin(\object{X})}$. For any multiset $m$, we can then show that $w_{m}=[m].R^{!}_{m}$, where $[m]$ denotes the multinomial coefficient $[m]=(\sum_{x\in\object{X}} m(x))!/ \prod_{x\in\object{X}} m(i)!$ (identifying $[m]$ as a function $\object{X}\rightarrow \naturalN$).
Although not equal, the two interpretations of exponentials are therefore related. Of course, the exact definition of Danos and Ehrhard can be considered in our setting by defining a variant of the exponential in which weights are scaled by a $\frac{1}{[m]}$ factor. This is however a less general construction as it can only be performed when the latter scalar can be defined.

\section{Perspectives}

%We defined the notion of \emph{triskell} in a category $\cat$, and then defined the category of triskells over a fixed monoid object $\object{\Omega}$ in $\cat$ following the definition of the category of spans in $\cat$. This category of triskells both abstracts the constructions of the author's interaction graphs \cite{seiller-goim,seiller-goiadd,seiller-goig} and generealises the category of weighted relations of Laird \emph{et al.} \cite{quantdenot}. Therefore triskells can be used to define both (quantitative) dynamic and (quantitative) denotational models of linear logic, though the tensor product is interpreted by distinct monoidal products. 

%This lead us to define an endofunctor on the category of triskells which mimics the antisymmetric Fock functor of linear algebra. This endofunctor maps the interpretation of a \MLL proof in dynamic semantics to the interpretation of the same proof in the denotational model. Moreover, this result is shown to hold in the general setting of quantitative models, and not only for the qualitative models. 

%Lastly, we discussed how this endofunctor not only maps the interpretation of proofs, but also maps the orthogonality relations used for constructing models of \MLL through double-glueing constructions. This implies that the interaction graphs models introduced by the author \cite{seiller-goiadd} are mapped to some quantitative generalisation of Girard's \emph{coherence spaces}. We ended the paper defining this notion of \emph{quantitative coherence spaces}, and showing how they provide a model for linear logic.

It appears that the category of triskells provides an interesting generalisation to the category of weighted relations when it comes to defining quantitative denotational models. In this paper, we have presented the structure only for multiplicative linear logic, and it thus makes sense to study extensions to interpret proofs of larger fragments, e.g. additives and exponential connectives of linear logic, fixed points. As such extensions were considered by the author for the interaction graphs models, a related question is what these extensions are mapped to through the \enquote{functor} $\liftedfock$. 
Furthermore, these extensions can be studied not only at the level of proof interpretations, but also with respect to the double-glueing constructions. Indeed, quantitative coherence spaces models were defined on the category of triskells for larger fragments in the last section. It should be noted, however, that to define satisfying exponential connectives, one has to consider generalisations of graphs \cite{seiller-goig,seiller-goif}. 

 %These coherence spaces models are related to the interaction graphs models. 

Lastly, the notion of triskell seems of interest in itself, especially considering the fact that its definition is internal to the underlying category $\cat$. It then appears that both dynamic and denotational models can be defined in categories of triskells over any topos instead of a category of sets.

\clearpage
\bibliographystyle{plainurl}
%\bibliographystyle{../../Styles/LNCS/splncs03}
%\bibliography{../../Common/thomas}

\begin{thebibliography}{22}

\bibitem{baillot-phd}
Patrick Baillot.
\newblock {\em Approches dynamiques de la logique lin\'{e}aire: jeux et
  g\'{e}om\'{e}trie de l'interaction}.
\newblock PhD thesis, Aix-Marseille 2 University, January 1999.

\bibitem{spans}
Jean B\'{e}nabou.
\newblock Introduction to bicategories.
\newblock In {\em Reports of the Midwest Category Seminar}, {\em
  Lecture Notes in Mathematics} 47, pages 1--77. Springer, 1967.
\newblock \href {http://dx.doi.org/10.1007/BFb0074299}
  {\path{doi:10.1007/BFb0074299}}.

\bibitem{CalderonMcCusker}
Ana~C Calderon and Guy~A McCusker.
\newblock Understanding game semantics through coherence spaces.
\newblock  In {\em MFPS 2010}.

\bibitem{probcoh}
Vincent Danos and Thomas Ehrhard.
\newblock Probabilistic coherence spaces as a model of higher-order
  probabilistic computation.
\newblock {\em Information and Computation}, 209, 2011.

\bibitem{FKdet}
Bent Fuglede and Richard~V. Kadison.
\newblock Determinant theory in finite factors.
\newblock {\em Annals of Mathematics}, 56(2), 1952.

\bibitem{linearlogic}
Jean-Yves Girard.
\newblock Linear logic.
\newblock {\em Theoretical Computer Science}, 50(1):1--102, 1987.

\bibitem{FoncteurAnalytiques}
Jean-Yves Girard.
\newblock Normal functors, power series and $\lambda$-calculus.
\newblock {\em Annals of Pure and Applied Logic}, 37(2):129--177, 1988.
\newblock \href {http://dx.doi.org/10.1016/0168-0072(88)90025-5}
  {\path{doi:10.1016/0168-0072(88)90025-5}}.

\bibitem{qcs}
Jean-Yves Girard.
\newblock {\em Between logic and quantic : a tract}, pages 346--381.
\newblock Number 316 in London Mathematical Society Lecture Note Series.
  Cambridge University Press, 2004.

\bibitem{goi5}
Jean-Yves Girard.
\newblock Geometry of interaction $\text{V}$: Logic in the hyperfinite factor.
\newblock {\em Theoretical Computer Science}, 412:1860--1883, 2011.

\bibitem{catgoi}
Esfandiar Haghverdi and Philip Scott.
\newblock A categorical model for the geometry of interaction.
\newblock {\em Theoretical Computer Science}, 350(2):252--274, 2006.

\bibitem{haghverdiscottdenot}
Esfandiar Haghverdi and Philip~J. Scott.
\newblock From geometry of interaction to denotational semantics.
\newblock {\em Electronic Notes in Theoretical Computer Science}, 122:67--87,
  2005.

\bibitem{doubleglueing}
Martin Hyland and Andrea Schalk.
\newblock Glueing and orthogonality for models of linear logic.
\newblock {\em Theoretical Computer Science}, 294, 2003.
\newblock \href {http://dx.doi.org/10.1016/S0304-3975(01)00241-9}
  {\path{doi:10.1016/S0304-3975(01)00241-9}}.

\bibitem{tmc}
Andr\'{e} Joyal, Ross Street, and Dominic Verity.
\newblock Traced monoidal categories.
\newblock {\em Mathematical Proceedings of the Cambridge Philosophical
  Society}, 119:447--468, 1996.

\bibitem{quantdenot}
Jim Laird, Guy McCusker, Giulio Manzonetto, and Michele Pagani.
\newblock Weighted relational models of typed lambda-calculi.
\newblock In {\em IEEE/ACM Logic in Computer Science (LICS) 2013}.

\bibitem{lang}
Serge Lang.
\newblock {\em Algebra}.
\newblock Number 211 in Graduate Texts in Mathematics. Springer Verlag, 3rd
  edition, 2002.

\bibitem{ScottOutline}
Dana Scott.
\newblock Outline of a mathematical theory of computation.
\newblock Technical Report PRG02, OUCL, November 1970.

\bibitem{seiller-goim}
Thomas Seiller.
\newblock Interaction graphs: Multiplicatives.
\newblock {\em Annals of Pure and Applied Logic}, 163:1808--1837, December
  2012.
\newblock \href {http://dx.doi.org/10.1016/j.apal.2012.04.005}
  {\path{doi:10.1016/j.apal.2012.04.005}}.

\bibitem{seiller-masas}
Thomas Seiller.
\newblock A correspondence between maximal abelian subalgebras and linear
  logic fragments.
\newblock {\em Mathematical Structures in Computer Science}, to appear, 2016.

\bibitem{seiller-goiadd}
Thomas Seiller.
\newblock Interaction graphs: Additives.
\newblock {\em Annals of Pure and Applied Logic}, 167:95 -- 154, 2016.
\newblock \href {http://dx.doi.org/10.1016/j.apal.2015.10.001}
  {\path{doi:10.1016/j.apal.2015.10.001}}.

\bibitem{seiller-goif}
Thomas Seiller.
\newblock Interaction graphs: Full linear logic.
\newblock In {\em IEEE/ACM Logic in Computer Science (LICS) 2016}.

\bibitem{seiller-goig}
Thomas Seiller.
\newblock Interaction graphs: Graphings.
\newblock {\em Annals of Pure and Applied Logic}, 2016.
\newblock to appear.
\newblock URL: \url{http://arxiv.org/pdf/1405.6331}.

\bibitem{detsemiring}
Yi-Jia Tan.
\newblock Determinants of matrices over semirings.
\newblock {\em Linear and Multilinear Algebra}, 62(4):498--517, 2014.
\newblock \href {http://dx.doi.org/doi:10.1080/03081087.2013.784285}
  {\path{doi:doi:10.1080/03081087.2013.784285}}.

\end{thebibliography}

\end{document}